\documentclass[prx,preprint]{revtex4}
\usepackage{graphicx}
\usepackage{dcolumn}
\usepackage{bm}
\usepackage{color}
\usepackage{soul}

\newcommand{\nc}{\newcommand}
\nc{\I}{$I$}
\nc{\II}{$II$}
\def\beq{\begin{equation}}
\def\eeq{\end{equation}}
\nc{\III}{$III$}
\nc{\nn}{\nonumber}
\def\e{\mathcal{E}}
\def\U{\mathcal{U}}

\nc{\XYZ }{}
\nc{\ABC }{\st}

\usepackage{xr}
\makeatletter
\newcommand*{\addFileDependency}[1]{
  \typeout{(#1)}
  \@addtofilelist{#1}
  \IfFileExists{#1}{}{\typeout{No file #1.}}
}
\makeatother

\newcommand*{\myexternaldocument}[1]{%
    \externaldocument{#1}%
    \addFileDependency{#1.tex}%
    \addFileDependency{#1.aux}%
}
\myexternaldocument{SI-rev1}

\begin{document}
\newcommand{\av}[1]{\langle #1 \rangle}

\title{``Hot'' electrons in metallic nanostructures - non-thermal carriers or heating?}

\author{Yonatan Dubi}
\affiliation{Department of Chemistry and the Ilse Katz Center for nanoscale Science and Technology, Ben-Gurion University of the Negev, Beer Sheva, Israel}
\email{jdubi@bgu.ac.il}

\author{Yonatan Sivan}
\affiliation{Unit of Electro-Optics Engineering and the Ilse Katz Center for nanoscale Science and Technology, Ben-Gurion University of the Negev, Beer Sheva, Israel}
\email{sivanyon@bgu.ac.il}

\date{\today}


\begin{abstract}
Understanding the interplay between illumination and the electron distribution in metallic nanostructures is a crucial step towards developing applications such as plasmonic photo-catalysis for green fuels, nano-scale photo-detection and more. Elucidating this interplay is challenging, as it requires taking into account all channels of energy flow in the electronic system. Here, we develop such a theory, which is based on a coupled Boltzmann-heat equations and requires only energy conservation and basic thermodynamics, where the electron distribution, and the electron and phonon (lattice) temperatures are determined {\em uniquely}. Applying this theory to realistic illuminated nanoparticle systems, we find that the electron and phonon temperatures are similar, thus justifying the (classical) single temperature models. We show that {\XYZ while the fraction of high-energy ``hot'' carriers compared to thermalized carriers grows substantially with illumination intensity, it remains extremely small (on the order of $10^{-8}$). Importantly,}  most of the absorbed illumination power goes into heating rather than generating hot carriers, thus rendering plasmonic hot carrier generation extremely inefficient. Our formulation allows for the first time a unique quantitative comparison of theory and measurements of steady-state electron distributions in metallic nanostructures.
\end{abstract}

\maketitle



What happens to electrons in a metal when they are illuminated? This fundamental problem is a driving force in shaping modern physics since the discovery of the photo-electric effect. In recent decades, this problem resurfaced from a new angle, owing to developments in the field of nano-plasmonics~\cite{plasmonics_review_Brongersma_2010,plasmonics_review_nat_phot_2012}, where metallic nanostructures give rise to resonantly enhanced local electromagnetic fields, and hence, to controllable optical properties.

Even more recently, there is growing interest in controlling also the electronic and chemical properties of metal nanostructures. In particular, upon photon absorption, energy is transferred to the electrons in the metal, thus driving the electron distribution out of equilibrium; The generated non-thermal electrons - sometimes (ill)referred to as ``hot'' electrons - can be exploited for photo-detection~\cite{Uriel_Schottky,Uriel_Schottky2,Valentine_hot_e_review} and up-conversion~\cite{Guru_APL_up_conversion,Guru_APL_up_conversion_exp}. Many other studies claimed that ``hot'' electrons can be exploited in photo-catalysis, namely, to drive a chemical reaction such as hydrogen dissociation, water splitting~\cite{chem_rev_photochemistry_2006,hot_e_review_Purdue,plasmonic_photocatalysis_1,plasmonic_photocatalysis_Clavero,plasmonic-chemistry-Baffou,hot_es_review_2015_Moskovits,Aruda4212} or artificial photosynthesis~\cite{Moskovits_hot_es,plasmonic_photo_synthesis_Misawa}; These processes have an immense importance in paving the way towards realistic alternatives for fossil fuels.

Motivated by the large and impressive body of experimental demonstrations of the above-mentioned applications, many theoretical studies address the question: how many non-equilibrium high energy (``hot'') electrons are generated for a given illumination. Na\"ively, one would think that the answer is already well-known, but in fact, finding a quantitative answer to this question is a challenging task. A complete theory of non-equilibrium carrier generation should not only include a detailed account of the non-equilibrium nature of the electron distribution, but also account for the possibility of the electron temperature to increase (via $e-e$ collisions), the phonon temperature to increase (due to $e-ph$ collisions), as well as for energy to leak from the lattice to the environment (e.g., a substrate or solution). The model should then be used for finding the steady-state non-equilibrium electron distribution which is established under {\em continuous wave} (CW) illumination, as appropriate for technologically-important applications such as photodetection and photo-catalysis.

Quite surprisingly, to date, there is no comprehensive theoretical approach that takes all these elements into account. Typically, the transient electron dynamics is studied~\cite{non_eq_model_Lagendijk,delFatti_nonequilib_2000,vallee_nonequilib_2003,Italians_hot_es,Govorov_nature_nano,GdA_hot_es}, focusing on an accurate description of the material properties, e.g., metal band structure and collision rates~\cite{Atwater_Nat_Comm_2014,Louie_photocatalysis,Brown_PRL_2017}; some studies also accounted 
for the electron temperature dynamics~\cite{Italians_hot_es,GdA_hot_es} and (to some extent) for the permittivity~\cite{GdA_hot_es} dynamics. 
On the other hand, the few pioneering theoretical studies of the steady-state non-equilibrium under CW illumination~\cite{Govorov_1,Govorov_2} accounted for the electron distribution in great detail, 
but assumed that the electron and phonon (lattice) temperature are both at room temperature. In~\cite{Govorov_ACS_phot_2017} an ``effective'' electron temperature is referred to~\footnote{No formal definition for it is given in~\cite{Govorov_ACS_phot_2017}. }; it is assumed to be higher than the environment temperature, but is pre-determined (rather than evaluated self-consistently). As discussed in SI Section~\ref{app:practical}, the chosen values for that ``effective'' electron temperature are questionable.



The fact that the phonon and electron temperatures were not calculated in previous theoretical studies of the ``hot electrons'' distribution is not a coincidence. After all, the system is out of equilibrium, so how can one define a unique value for the temperature, inherently an equilibrium property?~\cite{Puglisi}. Yet, it is well-known that the temperature of metallic nanostructures {\em does} increase upon CW illumination, sometime to the degree of melting (or killing cancer cells); this process is traditionally described using classical, single temperature heat equations (see, e.g.,~\cite{baffou2013thermo,Two_temp_model,Abajo_nano-oven}).

Here, we suggest a unique self-contained theory for the photo-generation of non-equilibrium energetic carriers in metal nanostructures that reconciles this ``paradox''. 
The framework we chose is the quantum-like version of the Boltzmann equation (BE), which is in regular use for describing electron dynamics in metallic systems more than a few nm in size~\cite{Ziman-book,Ashcroft-Mermin,Quantum-Liquid,Dressel-Gruner-book,non_eq_model_Lagendijk,delFatti_nonequilib_2000,non_eq_model_Rethfeld,vallee_nonequilib_2003,Italians_hot_es,GdA_hot_es,Seidman-Nitzan-non-thermal-population-model}. We employ the relaxation time approximation for the electron-electron thermalization channel to determine the electron temperature without ambiguity. Furthermore, on top of the BE we add an equation for the phonon temperature such that together with the integral version of the BE, our model equations provide a microscopic derivation of the extended two temperature model~\cite{Two_temp_model,Abajo_nano-oven}. In particular, the electron and phonon temperatures are allowed to rise above the ambient temperature and energy can leak to the environment while energy is conserved in the photon-electron-phonon-environment system. These aspects distinguish our calculation of the steady-state non-equilibrium from previous ones~\footnote{Govorov and Besteiro claim in~\cite{Govorov_Comment_ArXiv} that $T_{ph} > T_{env}$ in all their calculations; however, there is no indication for this in their published papers. }.

Using our theory, we show that the population of non-equilibrium energetic electrons and holes~\footnote{Negative values of the deviation from thermal equilibrium ($f - f^T$, see below) are referred to as holes, regardless of their position with respect to the Fermi energy. This nomenclature is conventional within the literature~\cite{Quantum-Liquid}. } can increase dramatically under illumination, yet this process is {\em extremely inefficient}, as almost all the absorbed energy leads to heating; the electron and phonon temperatures are found to be essentially similar, thus justifying the use of the classical single temperate heat  model~\cite{thermo-plasmonics-review}. Somewhat surprisingly, we find that just above (below) the Fermi energy, the non-equilibrium consists of holes (electrons), rather than the other way around; we show that this behaviour is due to the dominance of $e-ph$ collisions. All these results are very different from those known for electron dynamics under ultrafast illumination, as well as from previous studies of the steady-state scenario that did not account for all three energy channels (e.g.,~\cite{Manjavacas_Nordlander,Govorov_1,Govorov_2,Govorov_ACS_phot_2017}). Detailed comparison to earlier work is presented throughout the main text and the supplementary information (SI).
\newline
\vskip 0.25truecm
{\large \bf{Model}}\newline
We start by writing down the Boltzmann equation in its generic form,
\begin{eqnarray}\label{eq:BE}
 \frac{\partial f\left(\mathcal{E},T_e,T_{ph}\right)}{\partial t}
= \left(\frac{\partial f}{\partial t}\right)_{ex} + \left(\frac{\partial f}{\partial t}\right)_{e-e} + \left(\frac{\partial f}{\partial t}\right)_{e-ph}.
\end{eqnarray}
Here, $f$ is the electron distribution function at an energy $\e$, electron temperature $T_e$ and phonon temperature $T_{ph}$~\footnote{We neglect the deviation of the phonon system from thermal equilibrium. This is an assumption that was adopted in almost all previous studies on the topic; accounting for the phonon non-equilibrium can be done in a similar way to our treatment of the electron non-equilibrium, see e.g.,~\cite{Italians_hot_es,Baranov_Kabanov_2014}. }, representing the population probability of electrons in a system characterized by a continuum of states within the conduction band; finding it for electrons under (CW) illumination is our central objective.

\begin{figure}[h!]
\centering{\includegraphics[width=13 cm]{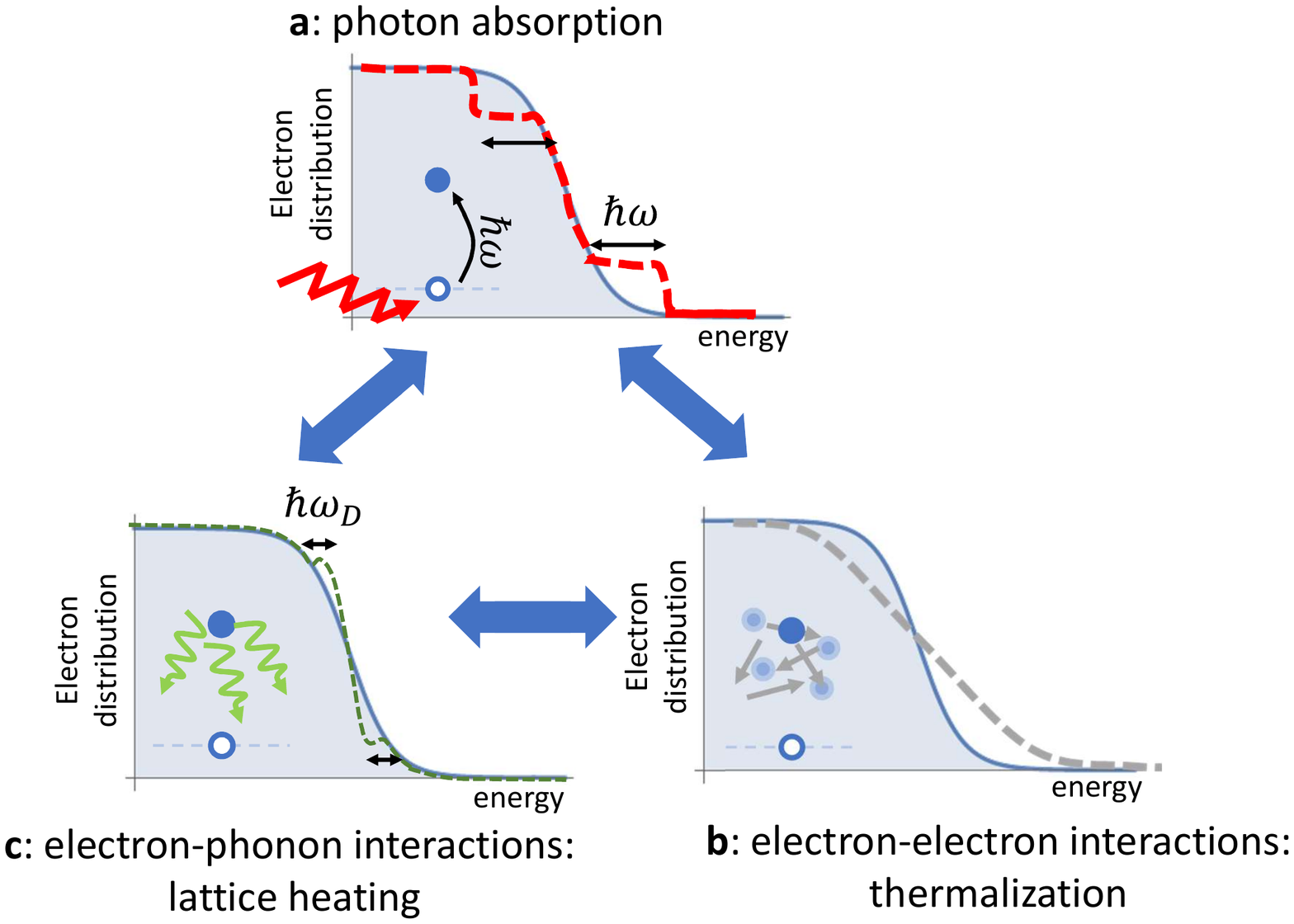}
\includegraphics[width=8.5cm]{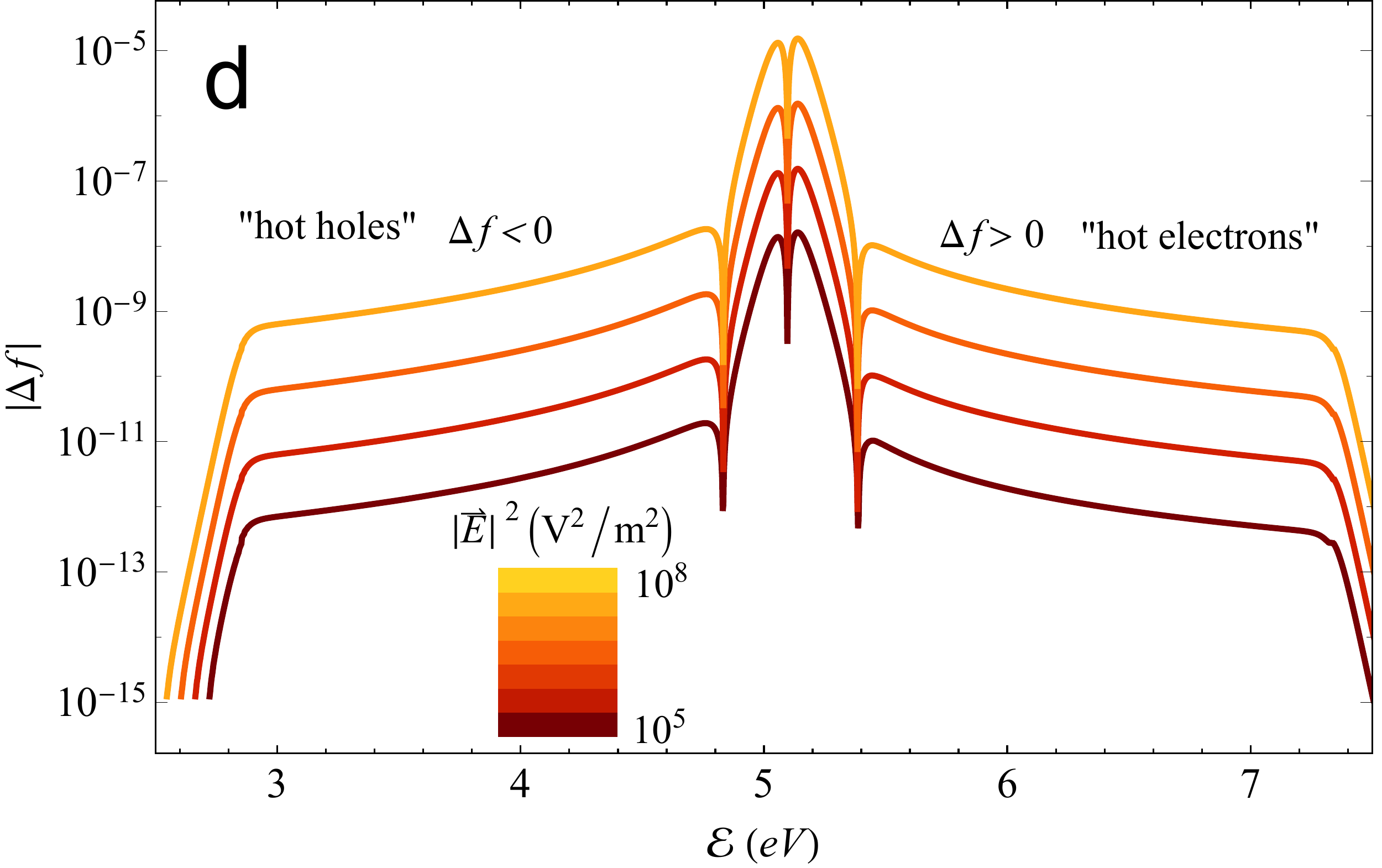}}
\caption{{\bf Full non-equilibrium electron distribution under illumination}. The steady-state of the system is determined by the balance of three processes, shown on the background of the thermal distribution (Grey). {\bf a}: absorption of photons by an electron, with an energy quanta $\hbar \omega$. {\bf b}: electron (red) - phonon (green) scattering, which leads to lattice heating.{\bf c}: electron-electron scattering, which leads to thermalization and electron heating. In addition, the excess thermal energy from the lattice can be transferred to the environment. {\bf d}: Deviation from the equilibrium distribution at the ambient electron temperature, namely, $\Delta f \equiv f(\e,T_e,T_{ph}) - f^T(\e,T_{env})$, as a function of electron energy for various incoming field levels; the system is a bulk Ag illuminated by $\hbar \omega = 2.25$eV photons, see all parameters values in Table~\ref{app:practical}. Non-thermal hole densities, which correspond to $\Delta f < 0$, are shown for simplicity in opposite sign. The dashed vertical line represents the Fermi energy. The various dips are artifacts of the semilogarithmic scale - they represent sign changes of $\Delta f$.}
\label{fig:dist-f-f_T_env}\end{figure}

The right-hand side of the BE describes three central processes which determine the electron distribution. 
Electron excitation due to photon absorption increases the electron energy by $\hbar \omega$, thus, generating an electron and a hole, see Fig.~\ref{fig:dist-f-f_T_env}(a) and Fig.~\ref{fig:BE_terms}
(a); it is described (via the term $\left(\frac{\partial f}{\partial t}\right)_{ex}$) using an improved version of the Fermi golden rule type form suggested in~\cite{delFatti_nonequilib_2000,vallee_nonequilib_2003,Seidman-Nitzan-non-thermal-population-model,GdA_hot_es} which here also incorporates explicitly the absorption lineshape of the nanostructure, see Eq.~(\ref{eq:QM-like-excitation}). 

Electron-phonon ($e-ph$) collisions cause energy transfer between the electrons and lattice; they occur within a (narrow) energy window (whose width is comparable to the Debye energy) near the Fermi energy, see Fig.~\ref{fig:dist-f-f_T_env}(b) and Fig.~\ref{fig:BE_terms}
(b). They are described using a general Bloch-Boltzmann-Peierls form~\cite{Ziman-book,PB_Allen_e_ph_scattering,delFatti_nonequilib_2000}.

Electron-electron ($e-e$) collisions 
lead to thermalization. They occur throughout the conduction band, but are strongly dependent on the energy - for carrier energies close to the Fermi energy they are relatively slower than for electrons with energies much higher than the Fermi energy, which can be as fast as a few tens of femtoseconds, see Fig.~\ref{fig:dist-f-f_T_env}(c) and~\cite{Italians_hot_es,hot_es_Atwater,GdA_hot_es}. Traditionally, two generic models are used to describe $e-e$ collisions. The exact approach invokes the 4-body interactions between the incoming and outgoing particles within the Fermi golden rule formulation, see e.g.,~\cite{Ziman-book,PB_Allen_e_ph_scattering,non_eq_model_Lagendijk,Dressel-Gruner-book,delFatti_nonequilib_2000,vallee_nonequilib_2003,Italians_hot_es}. This approach has two main drawbacks - first, evaluation of the resulting collision integrals is highly time-consuming~\cite{vallee_nonequilib_2003}; second, it is not clear what is the state into which the system wishes to relax (although it is clear that it should flow into a Fermi distribution at equilibrium conditions). 

A popular alternative is to adopt the so-called relaxation time approximation, whereby it is assumed that the non-equilibrium electron distribution relaxes to a Fermi-Dirac form $f^T(\e,T_e)$ ~\cite{Ziman-book,non_eq_model_Lagendijk,Dressel-Gruner-book,Govorov_ACS_phot_2017} with a well-defined temperature $T_e$, namely, $\left(\frac{\partial f(\e)}{\partial t}\right)_{e-e} = - \frac{f - f^T(T_e)}{\tau_{e-e}(\e)}$, where $\tau_{e-e}(\e)$ is the electron collision time. The electron temperature that characterizes that Fermi-Dirac distribution is the temperature that the electron subsystem will reach if the illumination is stopped and no additional energy is exchanged with the phonon subsystem. The relaxation time approximation is known to be an excellent approximation for small deviations from equilibrium (especially assuming the collisions are elastic and isotropic~\cite{Lundstrom-book}). In this approach, $e-e$ collision integral is simple to compute, and the physical principle which is hidden in the full collision integral description, namely, the desire of the electron system to reach a Fermi-Dirac distribution, is illustrated explicitly. Most importantly, the relaxation time approximation allows us to eliminate the ambiguity in the determination of the temperature of the electron subsystem. The collision time itself $\tau_{e-e}(\e)$ is evaluated by fitting the standard expression from Fermi-Liquid Theory to the computational data of Ref.~\cite{GdA_hot_es}, see SI Section~\ref{app:tau_ee}.


What remains to be done is to determine $T_{ph}$ - it controls the rate of energy transfer from the electron subsystem to the phonon subsystem, and then to the environment. Recent studies of the steady-state non-equilibrium in metals (e.g.,~\cite{Govorov_1,Govorov_2,Govorov_ACS_phot_2017})
relied on a fixed value for $T_{ph}$ (choosing it to be either identical to the electron temperature, or to the environment temperature~\footnote{see footnote [56]. }) and/or treated the rate of $e-ph$ energy transfer using the relaxation time approximation with a $e-ph$ collision rate which is independent of the field and particle shape. 
While these approaches ensure that energy is conserved in the electron subsystem, they ignore the dependence of the energy transfer to the environment on the nanoparticle shape, the thermal properties of the host material, the electric field strength and the temperature difference. Therefore, not only these phenomenological approaches fail to ensure energy conservation in the complete system (photons, electrons, phonons and environment), but they also fail to provide a correct quantitative prediction of the electron distribution near the Fermi energy (which is strongly dependent on $T_{ph}$) and provides incorrect predictions regarding the role of nanoparticle shape and host properties on the steady-state electron distribution and the temperatures (see further discussion in~\cite{Dubi-Sivan-Faraday}).


In order to determine $T_{ph}$ self-consistently while ensuring energy conservation, one has to account for the ``macroscopic'' properties of the problem. Specifically, 
we multiply Eq.~(\ref{eq:BE}) by the product of the electron energy $\e$ and the density of electron states $\rho_e(\e)$ and integrate over the electron energy. The resulting equation describes the dynamics of the energy of the 
electrons, 
\begin{equation}\label{eq:U_e}
\frac{d \U_e
}{dt} = W_{ex} - W_{e-ph}. 
\end{equation}
Eq.~(\ref{eq:U_e}) has a simple and intuitive interpretation: the dynamics of the electron energy 
is determined by the balance between the energy that flows in due to photo-excitation ($W_{ex} \equiv \int \e \rho_e(\e) \left(\frac{\partial f}{\partial t}\right)_{ex} d\e$) 
and the energy that {\em flows out} to the lattice ($W_{e-ph}
\equiv - \int \e \rho_e(\e) \left(\frac{\partial f
}{\partial t}\right)_{e-ph} d\e$, see SI Section~\ref{app:e-ph}). 

In similarity to Eq.~(\ref{eq:U_e}), the total energy of the lattice, $\mathcal{U}_{ph}$, is balanced by the heat flowing in from the electronic system and flowing out to the environment, namely,
\begin{equation}
\frac{d \U_{ph}}{dt}
= W_{e-ph} - G_{ph-env}(T_{ph} - T_{env}). \label{Eq:U_ph_eq}
\end{equation}
Here, 
$T_{env}$ is the temperature of the environment far from the nanostructure and $G_{ph-env}$ is proportional to the thermal conductivity of the environment; it is strongly dependent on the nanostructure geometry (e.g., exhibiting inverse proportionality to the particle surface area for spheres). 


Eqs.~(\ref{eq:BE})-(\ref{Eq:U_ph_eq}) provide a general formulation for the non-thermal electron generation, electron temperature and lattice temperature in metal nanostructures under arbitrary illumination conditions, see also discussion in SI Section~\ref{app:practical}. Once a steady-state solution for these equations is found, energy conservation is ensured - the power flowing into the metal due to photon absorption is exactly balanced by heat leakage to the environment. Within the relaxation time approach, there is only one pair of values for the electron and phonon temperatures for which this happens. Our ``macroscopic'' approach thus allowed us to determine the temperatures in a system which is out of equilibrium in a  unique and unambiguous way. 


The equations require as input the local electric field distribution from a solution of Maxwell's equations for the nanostructure of choice, see SI Section~\ref{app:practical}. In what follows, we numerically search for the steady-state ($\partial/\partial t = 0$) solution of these (nonlinear) equations for the generic (and application-relevant) case of CW illumination. For concreteness, we chose parameters for Ag, taken from comparison to experiments of ultrafast illumination~\cite{delFatti_nonequilib_2000}; the photon energy and local field values are chosen to coincide with the localized plasmon resonance of a Ag nano-sphere in a high permittivity dielectric, in similarity to many experiments~\cite{Moskovits_hot_es,Halas_dissociation_H2_TiO2}\footnote{In particular, the local field in this configuration gives a plasmonic near-field enhancement, of at least an order of magnitude, depending on the geometry and material quality. Our approach applies to any other configuration just by scaling the local field appropriately, see SI Section~\ref{app:practical}.}, see Table~\ref{app:practical}; this configuration also justifies the neglect of interband transitions (see discussion in SI Section~\ref{app:Boltzmann}) and field inhomogeneities (see discussion in SI Section~\ref{app:practical}). As we demonstrate, this generic case leads to several surprising qualitative {\em new} insights, as well as to quantitative predictions of non-equilibrium carrier distributions.


\vskip 0.5truecm
{\large \bf{Results}}\newline

{\bf{Electron distribution}}.~~Fig.~\ref{fig:dist-f-f_T_env}(d) shows the deviation of the electron distribution from the distribution at the ambient temperature (i.e., in the dark), $\Delta f \equiv f(\e,T_e,T_{ph}) - f^T(\e,T_{env})$, as a function of electron energy for various local field levels. The distributions depend on the local field quantitatively, but are qualitatively similar, showing that the resonant plasmonic near-field enhancement can indeed be used to increase the number of photo-generated ``hot'' electrons, as predicted and observed experimentally.

\begin{figure*}
\centering{
\includegraphics[width=15.5cm]{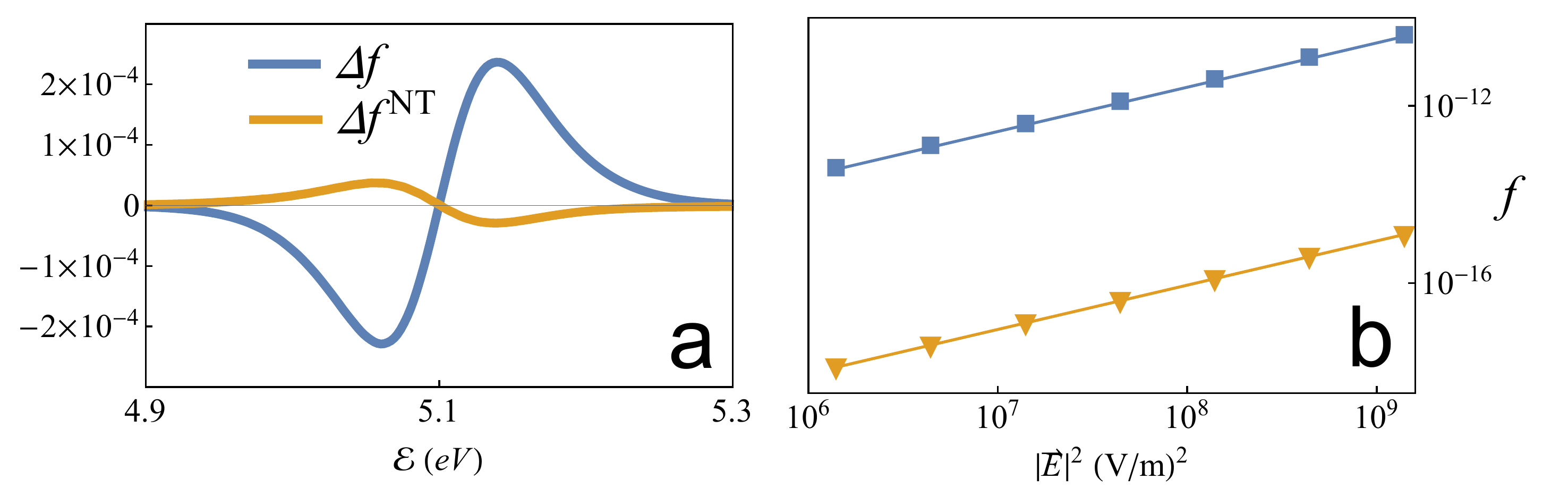}}
\caption{{\bf Non-thermal contribution to non-equilibrium}.
{\bf a:} Comparison of the true non-equilibrium distribution $\Delta f^{NT} \equiv f(\e,T_e,T_{ph}) - f^T(\e,T_e)$ with $\Delta f$ within the energy range close to the Fermi energy for $|\vec{E}|^2 = 10^9 [V/m]$. The true non-equilibrium is smaller and of opposite sign, indicating on the presence of non-thermal holes (electrons) above (below) the Fermi energy. {\bf b:} The populations $f(\e)$ of electrons at $\e = 1.8$ eV above the Fermi level (blue rectangles) and electrons at $\e = 2.5$ eV ($> \hbar \omega$) above the Fermi level (yellow triangles), all as a function of local field, showing a quadratic dependence between illumination field and ``hot'' carrier population (with a similar slope). } \label{fig:dist-f-f_Te}
\end{figure*}

The overall deviation from equilibrium (see scale in Fig.~1(d)) is minute, thus, justifying {\em a-posteriori} the use of the relaxation time approximation; in fact, near the Fermi energy, the deviation takes the regular thermal form, namely, it is identical to the population difference between two thermal distributions, thus justifying the assignment of the system with electron and phonon temperatures. In particular, the change of population is largest near the Fermi energy; specifically, $\Delta f > 0$ ($ < 0$) above (below) the Fermi energy, corresponding to electrons and holes, respectively~\footnote{We note that since $\Delta f$ (and $\Delta f^{NT}$ below) are not distributions, but rather, differences of distributions, they can attain negative numbers, representing holes.}, see Fig.~\ref{fig:dist-f-f_Te}(a). This is in accord with the approximate (semi-classical) solution of the Boltzmann equation (see e.g.,~\cite{Ziman-book,Dressel-Gruner-book}) and the standard interpretation of the non-equilibrium distribution (see e.g.,~\cite{Manjavacas_Nordlander,Munday_hot_es}).

{\bf{The ``true'' non-thermal distribution}}.~~It is clear that the distributions $\Delta f$ in Fig.~\ref{fig:dist-f-f_T_env} mix the two components of the electron distribution, namely, the thermal and non-thermal parts. To isolate the non-thermal contribution, one should consider the deviation of the electron distribution from the distribution at the {\em steady-state} temperature, $\Delta f^{NT} \equiv f(\e,T_e,T_{ph}) - f^T(\e,T_e)$. Simply put, this is the ``true'' non-thermal part of the steady-state electron distribution, loosely referred in the literature as the ``hot electron distribution''.

Since the differences between $f^T(\e,T_e)$ and $f^T(\e,T_{env})$ occur mostly around the Fermi energy, it is instructive to study $\Delta f^{NT}$ in two energy regimes. First, Fig.~\ref{fig:dist-f-f_Te}(a) shows that near the Fermi energy, the population change is now about an order of magnitude smaller and of the opposite sign (in comparison to $\Delta f$, Fig.~\ref{fig:dist-f-f_T_env}(d)). This is a somewhat surprising result, which means that the non-thermal distribution just above (below) the Fermi energy is characterized by the presence of non-thermal holes (electrons). This result could only be obtained when the explicit separation of the three energy channels are considered, allowing $T_e$ to increase above $T_{env}$. Notably, this is the exact opposite of the regular interpretation of the non-equilibrium distribution (as e.g., in Fig.~\ref{fig:dist-f-f_T_env}(d) and standard textbooks~\cite{Ziman-book,Dressel-Gruner-book}) which 
result from a calculaiton that does not account for the 
electron temperature rise. From the physical point of view, this change of sign originates from $e-ph$ collisions, as it has the same energy-dependence as the Bloch-Boltzmann-Peierls term, compare Fig.~\ref{fig:dist-f-f_Te}(b) with Fig.~\ref{fig:BE_terms}(b). 

Second, further away from the Fermi energy, $\hbar \omega$-wide (roughly symmetric) shoulders are observed on both sides of the Fermi energy (Fig.~\ref{fig:dist-f-f_T_env}(d)), corresponding to the generation of non-thermal holes ($\Delta f^{NT} < 0$) and non-thermal electrons ($\Delta f^{NT} > 0$). It is these high energy charge carriers that are referred to in the context of catalysis of chemical reactions.

For energies beyond $\hbar \omega$ from the Fermi energy, the non-thermal distribution is much lower, as it requires multiple photon absorption~\footnote{Observing the expected multiple step structure~\cite{Italians_hot_es}, is numerically very challenging for the steady-state case. }. 
This implies that in order to efficiently harvest the excess energy of the non-thermal electrons, one has to limit the harvested energy to processes that require an energy smaller than $\hbar \omega$.

The non-thermal electron distributions we obtained look similar to those obtained by calculations of the excitation rates due to photon absorption~\cite{Atwater_Nat_Comm_2014,Manjavacas_Nordlander,Munday_hot_es,Louie_photocatalysis}. However, as pointed out in~\cite{Atwater_Nat_Comm_2014,Govorov_ACS_phot_2017}, this approach yields the correct electron distribution only immediately after illumination by an ultrashort pulse (essentially before any scattering processes take place); this distribution would be qualitatively similar to the steady-state distribution only if all other terms in the BE were energy-independent, which is not the case (see SI Section~\ref{app:Boltzmann} and Fig.~\ref{fig:BE_terms}
). More specifically, this approach does not predict correctly the electron distribution near the Fermi energy; this means that the total energy stored in the electron system is not correctly accounted for and that the contribution of inter-band transitions to the non-equilibrium cannot be correctly determined. The main reason for these inaccuracies is that these studies did not correctly account for the electron and phonon temperatures, hence, the energy flow from the thermal electrons to the lattice such that quantitative conclusions on the distribution drawn in these studies should be taken with a grain of salt. Similar inaccuracies are found also in~\cite{Govorov_1,Govorov_2,Govorov_ACS_phot_2017,Govorov_nature_nano}.

On the other hand, these approaches can be used to provide a quantitative prediction of the electron distribution away from the Fermi energy, where $e-ph$ interactions are negligible (see~\cite[Section IIB]{Dubi-Sivan-Faraday}); peculiarly, however, this was not attempted previously~\cite{Atwater_Nat_Comm_2014,Manjavacas_Nordlander,Munday_hot_es,Louie_photocatalysis}, and instead, only claims about the qualitative features of the electron distributions were made. 

Our calculations also show that the number of photo-generated high energy electrons $\Delta f^{NT}$ is independent of $G_{ph-env}$ (see Fig.~\ref{fig:SM2} and discussion in SI Section~\ref{app:practical}). Since $G_{ph-env}$ is proportional to the thermal conductivity of the host and inversely proportional to the particle surface area, this implies that if a specific application relies on the number of high energy electrons, then, it will be relatively insensitive to the thermal properties of the host and the particle size
. Conversely, since the temperature rise is inversely proportional to $G_{ph-env}$ (see~\cite{thermo-plasmonics-basics} and Fig.~\ref{fig:SM2}), the difference in the photo-catalytic rate between the TiO$_2$ and SiO$_2$ substrates (compare~\cite{Halas_dissociation_H2_TiO2} and~\cite{Halas_H2_dissociation_SiO2}) is likely a result of a mere temperature rise, but is not likely to be related to the number of photo-generated high energy electrons (see further discussion in~\cite{Y2-eppur-si-riscalda}). 

\begin{figure}[h!]
\centering{
\includegraphics[width=8.5cm]{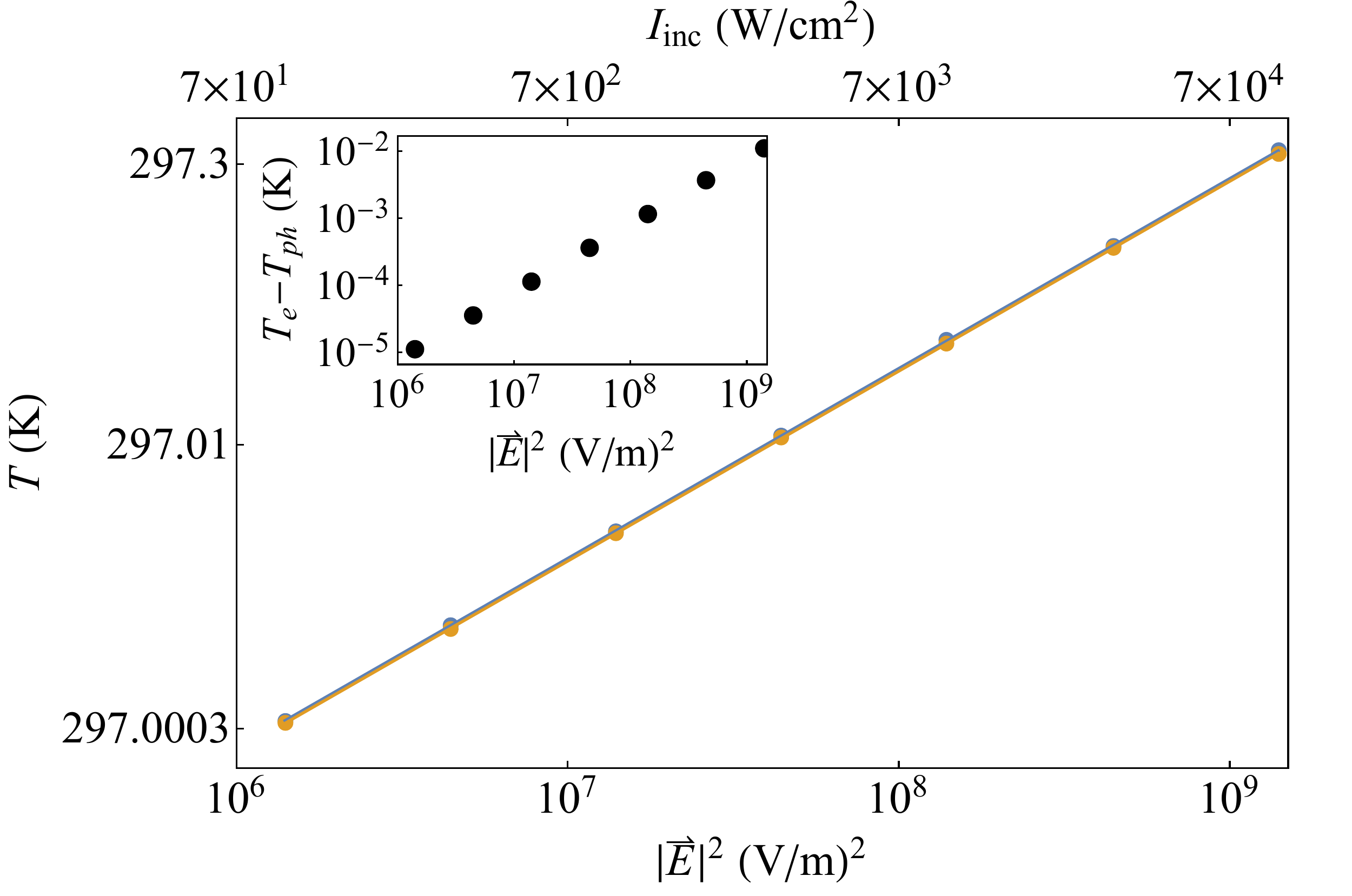}}
\caption{{\bf Temperature rise under illumination}. The electron (blue) and lattice (orange) temperatures extracted from the data of Figs.~\ref{fig:dist-f-f_T_env}-\ref{fig:dist-f-f_Te} as a function of the local field (loglog scale), showing a linear dependence on field-squared (top $x$-axis show the corresponding incident intensity). Inset: difference between electron and phonon temperatures as a function of field squared, showing that (i) the difference is also linear, and (ii) several orders of magnitude smaller than the temperatures themselves (making the electron and phonon temperatures essentially equal).} \label{fig:Ts}
\end{figure}

{\bf Electron and phonon temperatures.}~~As pointed above, our approach allows a quantitative estimate of both electron and phonon temperatures. In Fig.~\ref{fig:Ts}, these are plotted (on a log-log scale) as a function of the local field squared $|\vec{E}|^2$ (also translated into incident illumination intensity $I_{inc}$ in the upper x-axis for the specific case of a 5nm Ag sphere). As seen, both temperatures grow linearly with $|\vec{E}|^2$ over many decades of the field, as in the classical (single temperature) approach~\cite{thermo-plasmonics-review,Abajo_nano-oven}. In the inset we plot the difference between the electron and phonon temperatures as a function $|\vec{E}|^2$. This difference is also linear, and is seen to be much smaller (around two orders of magnitude) than the temperatures themselves.  
This is a nontrivial result, since the our non-equilibrium model equations exhibit an implicit nonlinear dependence on the temperatures. Fig.~\ref{fig:Ts} also shows that $T_e$ is only slightly higher than $T_{ph}$. 
This provides the first (qualitative and quantitative) justification, to the best of our knowledge, for the use of the single temperature heat equation in the context of metallic nanostructures under illumination~\cite{thermo-plasmonics-review, Abajo_nano-oven}; more generally, it provides a detailed understanding of the origin of the single-temperature model, as well as the limits to its validity (as at high intensities the electron-phonon temperature difference may become substantial).

\begin{figure}[h!]
\centering{\includegraphics[width=8.5cm]{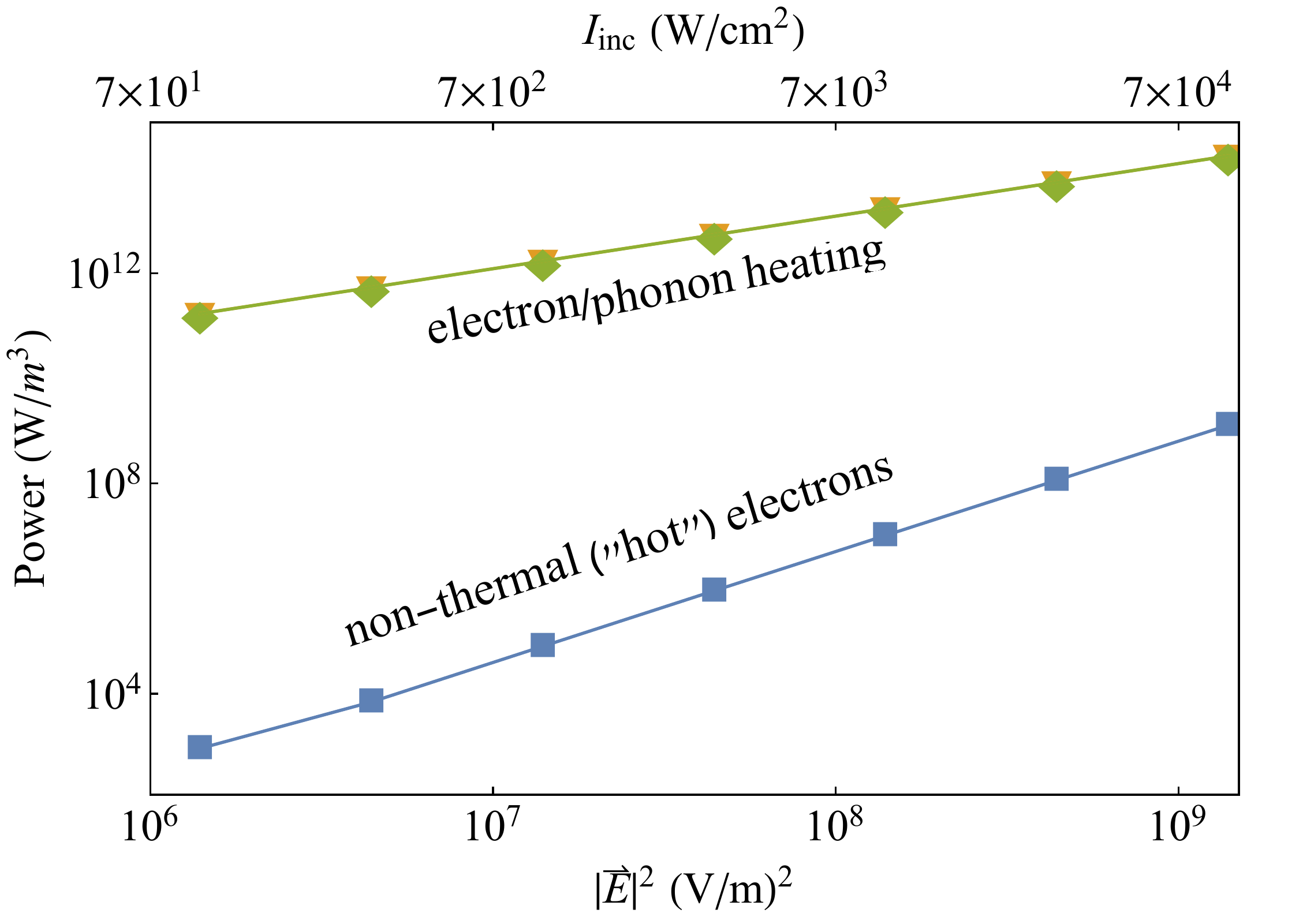}}
\caption{{\bf Power density and its distribution between the different channels.} Power densities going into the thermal electron and lattice systems ($W_{e-e}$ in green diamonds and $W_{e-ph}$ in orange triangles, respectively), compared with the power going to the non-thermal electrons ($W_{ex}^{NT}$ in blue squares), all as a function of the local field. The power fraction that flows into the thermal channels (i.e., to heat the systems) is substantially larger than that going into generating non-thermal electrons.  }
\label{fig:efficiency} \end{figure}

{\bf Efficiency.}~~Our approach allows us to deduce 
how the power density pumped into the metal by the absorbed photons splits into the non-thermal electrons and into heating the electrons and the phonons (see Fig.~\ref{fig:efficiency}), providing a way to evaluate the efficiency of the non-thermal electron generation (detailed calculation described in SI Section~\ref{sub:power}). Remarkably, one can see that the overall efficiency of the non-thermal electron generation is truly abysmal: At low intensities, the power channeled to the deviation from equilibrium ($W_{ex}^{NT} \equiv \int \e \rho_e(\e) \left(\frac{\partial f^{NT}}{\partial t}\right)_{ex} d\e$) is more than 8(!) orders of magnitude lower than the power invested in the heating of the electrons and phonons (which are accordingly nearly similar). This is in correlation with the results of Fig.~\ref{fig:dist-f-f_T_env}: most absorbed power leads to a change of the electron distribution near the Fermi energy, rather than to the generation of high energy electrons, as one might desire. This shows that any interpretation of experimental results which ignores electron and phonon heating should be taken with a grain of salt. It is thus the main result of the current study.

The performance of a ``hot'' electron system (say for catalysis or photo-detection, when electrons need to tunnel out of the nanoparticle) is essentially proportional to the electron distribution at the relevant energies (see SI Section~\ref{app:tunnel}). A comparison with the pure thermal distribution of high energy electrons (Fig.~\ref{fig:dist-f-f_Te}) shows that the absolute electron population can be many orders of magnitude higher compared to the thermal distribution at the steady-state temperature. Such an enhancement was indeed observed in ``hot'' electron based photodetection devices~\cite{Valentine_hot_e_review,Halas_hot_es_photodetection}, but not in ``hot'' electron photocatalysis~\cite{chem_rev_photochemistry_2006,hot_e_review_Purdue,plasmonic_photocatalysis_Linic,plasmonic_photocatalysis_Clavero,plasmonic-chemistry-Baffou,hot_es_review_2015_Moskovits,Govorov_nature_nano,Manjavacas_Nordlander}.

One can identify several pathways towards significant improvements of the efficiency of photo-generation of non-thermal electrons. In particular, as can be seen from Fig.~\ref{fig:efficiency}, as the local field is increased, the power fraction going to non-equilibrium increases to $10^{-5}$. This improvement motivates the study of the non-thermal electron distribution for higher intensities. Such study, however, will require extending the existing formulation by extracting self-consistently also the metal permittivity from the non-equilibrium electron distribution $f$ (like done above for the electron temperature). 
Other pathways for improved ``hot'' electron harvesting may rely on interband transitions due to photons with energies {\em far} above the interband threshold~\cite{Louie_photocatalysis,Alivisatos_Toste}, or optimizing the nanostructure geometry to minimize heating and maximize the local fields~\cite{Fabrizio_hot_es}, e.g., using
few nm particles (which support the same number of non-thermal carriers but lower heating levels).

Finally, the formulation we developed serves as an essential first step towards realistic calculations of the complete energy harvesting process, including especially the tunneling process, and the interaction with the environment, be it a solution, gas phase or a semiconductor. Our formulation enables a quantitative comparison with experimental studies of all the above processes and the related devices. Similarly, our formulation can be used to separate thermal and non-thermal effects in many other solid-state systems away from equilibrium, in particular, semiconductor-based photovoltaic and thermo-photovoltaic systems.



{\bf Acknowledgements.} The authors would like to thank G. Bartal, A. Govorov, J. Lischner, R. Oulton, A. Nitzan, S. Sarkar and I.W. Un for many fruitful discussions.

\appendix
\section{Supplementary Information: \\ ``Hot'' electrons in metallic nanostructures - non-thermal carriers or heating?}

\subsection{Solution of the quantum-like Boltzmann equation}\label{app:Boltzmann}

We determine the electron distribution in the conduction (sp) band, $f(\e,T_e,T_{ph})$, in a metal nanostructure under continuous wave (CW) illumination by solving the quantum-like Boltzmann equation (BE). This model is in wide use for such systems~\cite{Ziman-book,Ashcroft-Mermin,Lundstrom-book,non_eq_model_Lagendijk,delFatti_nonequilib_2000,vallee_nonequilib_2003,Italians_hot_es,Seidman-Nitzan-non-thermal-population-model}; It is valid for nanoparticles which are more than a few nm in size (hence, not requiring energy discretization)~\cite{GdA_hot_es,Govorov_ACS_phot_2017} and for systems where coherence and correlations between electrons are negligible. The latter assumption holds for a simple metal at room temperatures (or higher), as it has a large density of electrons and fast collision mechanisms. In order to include quantum finite size effects or quantum coherence effects, one can use the known relation between the discretized BE and quantum master equations~\cite{Goodnick_DM_BE,Chattah_DM_BE} or by replacing the BE by that equation~\cite{Govorov_1,Govorov_2,Govorov_ACS_phot_2017}.

For simplicity, we consider a quasi-free electron gas such that the conduction band is purely parabolic (with a Fermi energy of $\mathcal{E}_F = 5.1$eV and total size of $\mathcal{E}_{max} = 9$eV, typical to Ag, see~\cite{Ashcroft-Mermin}). This allows us to represent the electron states in terms of energy $\e$ rather than momentum. We also neglect interband ($d$ to $sp$) transitions - these have a small role when describing metals like Al 
illuminated by visible light, 
Ag for wavelengths longer than about $500$nm or so, 
or Au for near infrared frequencies, where a dominantly Drude response is exhibited. Furthermore, as noted in~\cite{Govorov_ACS_phot_2017}, interband transitions are not likely to generate electrons with energies far above the Fermi level unless the photon energy is much higher than the bandgap energy.

The resulting Boltzmann equation is
\begin{eqnarray}\label{eq:f_neq_dynamics}
\frac{\partial f\left(\mathcal{E}(\vec{k});T_e,T_{ph}\right)}{\partial t} &=& \underbrace{\left(\frac{\partial f}{\partial t}\right)_{ex}}_{photon\ absorption} + \underbrace{\left(\frac{\partial f}{\partial t}\right)_{e-ph}}_{e-ph\ collisions} + \underbrace{\left(\frac{\partial f}{\partial t}\right)_{e-e}}_{e-e\ collisions},
\end{eqnarray}
where $f$ is the electron distribution function at an energy $\e$, electron temperature $T_e$ and phonon temperature $T_{ph}$, representing the population probability of electrons in a system characterized by a continuum of states within the conduction band. The first term on the right-hand-side (RHS) of Eq.~(\ref{eq:f_neq_dynamics}) describes excitation of conduction electrons due to photon absorption, see SI Section~\ref{app:QM_absorption} below for its explicit form. The second term on the RHS of Eq.~(\ref{eq:f_neq_dynamics}) describes energy relaxation due to collisions between electrons and phonons, see SI Section~\ref{app:e-ph} below for its explicit form. This interaction makes the electrons in our model only quasi-free. The third term on the RHS of Eq.~(\ref{eq:f_neq_dynamics}) (see SI Section~\ref{app:tau_ee} below for its explicit form) represents the thermalization induced by $e-e$ collisions, i.e., the convergence of the {\em non}-thermal population into the {\em thermalized} Fermi-Dirac distribution, given by
\begin{equation}\label{eq:f_eq}
f^T\left(\mathcal{E};T_e\right) = \left(1 + e^{(\mathcal{E} - \e_F)/k_B T_e}\right)^{-1},
\end{equation}
where $k_B$ is the Boltzmann constant~\footnote{Note that we ignore here the difference between the Fermi energy and the chemical potential; we verified in simulations that the difference between them is truly negligible in all cases we studied. }. 

Note that our model does not require indicating what is the exact nature of the various collisions (Landau damping, surface/phonon-assisted, etc., see discussions in~\cite{hot_es_Atwater,Khurgin_Landau_damping,GdA_hot_es,Louie_photocatalysis}), but rather, it accounts only for their cumulative rate. Within this description, it was shown in~\cite{hot_es_Atwater,GdA_hot_es} that the total electron collision time is independent of the size of the metal nanoparticle~\footnote{Notably, this is in contrast to the claims in~\cite{hot_es_review_2015} which were not supported by evidence. }. 
Our model also 
does not account for electron acceleration due to the force exerted on them by the electric field (which involves a classical description, see SI Section~\ref{app:QM_absorption} below), nor for drift due to its gradients or due to temperature gradients; these effects will be small in the regime of intensities considered in our study, especially for few nm (spherical) particles (see also SI Section~\ref{app:practical} below)~\cite{thermo-plasmonics-basics,Un-Sivan-size-thermal-effect}.
Similar simplifications were adopted in most previous studies of this problem, e.g.,~\cite{Govorov_1,Govorov_2,Govorov_3,Manjavacas_Nordlander,GdA_hot_es,Govorov_ACS_phot_2017}). These neglected effects can be implemented in our formalism in a straightforward way.

\begin{figure}[h!]
\centering{\includegraphics[width=18cm,height=9cm]{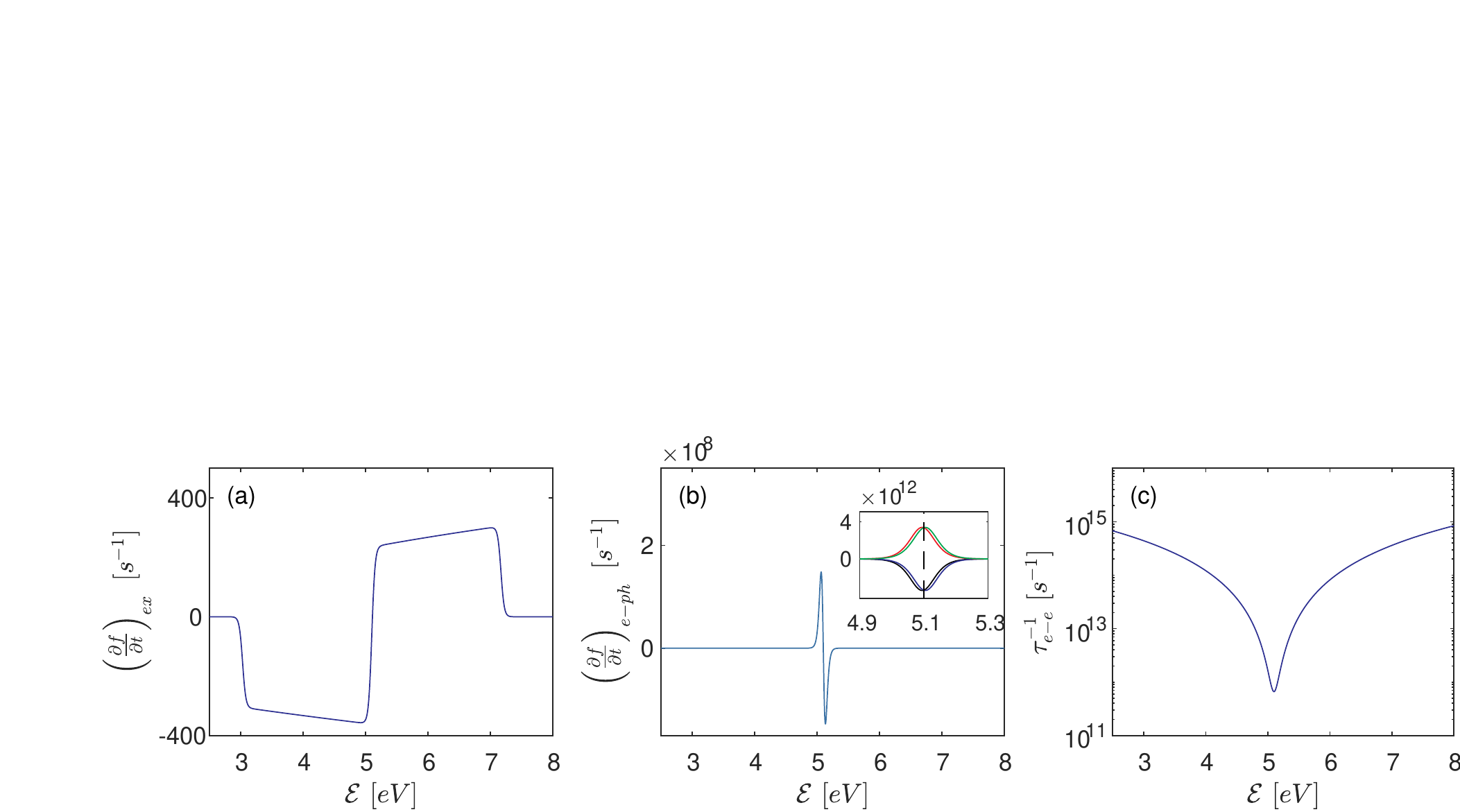}}
\caption{(Color online) (a) $\left(\frac{\partial f}{\partial t}\right)_{ex}$~(\ref{eq:QM-like-excitation}) as a function of electron energy for a local field of $|\vec{E}| = 7 \cdot 10^3 V/m$. (b) The $e-ph$ collision rate~(\ref{eq:e-ph-integral-term}) as a function of electron energy for $T_e - T_{ph} = 0.2^\circ$K. The inset shows the four competing phonon generation/absorption processes. (c) The $e-e$ collision rate as a function of electron energy, as given by Fermi liquid theory, Eq.~(\ref{eq:tau_ee_FLT}). } \label{fig:BE_terms}
\end{figure}

Finally, we emphasize that the results shown in the main text 
are not sensitive to the details of the general model. In fact, our procedure
can be made more system specific; for instance, 
the metal band structure can be taken into account~\cite{hot_es_Atwater,Louie_photocatalysis}, few nm nanoparticles can be studied by writing the BE in momentum space and discretizing it~\cite{GdA_hot_es}, and further quantum effects may be considered by replacing the BE by a quantum master equation~\cite{Govorov_1,Govorov_2,Govorov_ACS_phot_2017}. We do not expect any such change to have more than a moderate quantitative effect on the results shown below.

The steady-state solution of Eqs.~(\ref{eq:BE})-(\ref{Eq:U_ph_eq}) was attained numerically by writing the (thermal) electron and phonon energies as the product of the corresponding heat capacities and temperatures (see SI Section~\ref{sub:power}) and letting the system evolve naturally to the steady-state by ramping up slowly the electric field. Table~\ref{table:params} shows the values of all parameters used in our simulations. We observe that the results are insensitive to the initial conditions and choice of various parameter values.

\begin{table}\label{table:params}\centering
\caption{Parameters used in the simulations; values chosen for (low quality~\cite{Indian_Ag_ellipsometry_2014}) $5$nm Ag sphere. }

\begin{tabular}{|l|c|c|} \hline
parameter & parameter symbol & value \\
\hline
photon wavelength & $\lambda$ & $2.25$eV \\ \hline
metal permittivity & $\epsilon_{Ag}(\lambda)$ & $-8.5 + 1.8i$~\cite{Indian_Ag_ellipsometry_2014} \\ \hline
host permittivity & $\epsilon_h$ & $4.25$ \\ \hline
Fermi energy & $\e_F$ & $5.1$eV \\ \hline
conduction band width & $\e_{max}$ & $9$eV \\ \hline chemical potential & $\mu$ & $5.1$eV \\ \hline
ph-env coupling & $G_{ph-env}$ & $5 \cdot 10^{14} W/m^3 K$ \\ \hline
electron density & $n_e$ & $5.86 \cdot 10^{28} m^{-3}$ \\ \hline
speed of sound & $v_{ph}$ & $3650 $ m/s \\ \hline
environment temperature & $T_{env}$ & $297 K$ \\ \hline
electron mass & $m_e$ & $9.1 \cdot 10^{31}$ kg \\ \hline
\end{tabular}
\label{tab:params}
\end{table}

\subsubsection{The quantum mechanical excitation term}\label{app:QM_absorption}
Usually, the BE is regarded as a (semi-)classical model of electron dynamics. Indeed, several popular textbooks draw the links between the BE to the classical model of an electron motion in an electric field (e.g.,~\cite{Ziman-book,Ashcroft-Mermin,Dressel-Gruner-book,Marini_faraday_discuss_2019}). In this case, the change of momentum of the electrons (acceleration) due to the force exerted on them by the electric field corresponds to a {\em coherent} excitation term, i.e., a term which is proportional to $\frac{\partial}{\partial \mathcal{E}} \frac{\partial \mathcal{E}}{\partial \vec{k}} \cdot \frac{\partial \vec{k}}{\partial t} \sim \vec{v} \cdot \vec{E}$. However, since it relies on a classical field, this expression describes the photon-electron interaction correctly only if the energy imparted on the electron by the electric field is much greater than the energy of a single photon~\cite{non_eq_model_Rethfeld}. Since this is not the case, 
this term does not allow one to derive correctly the non-equilibrium distribution; in fact, this failure to produce experimental observations triggered Einstein to employ a quantized model for the photo-electric effect, and eventually led to the creation of quantum mechanics theory, as we know it.

In order to circumvent this problem {\em within} the BE, frequently the (semi-)classical (linear ($\sim \vec{E}$), coherent) excitation term is replaced by a quantum-like ($\sim |\vec{E}|^2$, incoherent) term derived from the Fermi golden rule~\cite{delFatti_nonequilib_2000,vallee_nonequilib_2003,Italians_hot_es,GdA_hot_es,Manjavacas_Nordlander,Seidman-Nitzan-non-thermal-population-model}. Early derivations of this term (e.g.~\cite{delFatti_nonequilib_2000}) did not supply a rigorous expression for its magnitude, but rather fit its magnitude to experimental results. Later studies attempted to link the magnitude of this term to the total absorbed power~\cite{Italians_hot_es}. A systematic derivation was provided in~\cite{GdA_hot_es}. 

Here, we employ the simpler, elegant expression proposed in~\cite{Seidman-Nitzan-non-thermal-population-model}, namely, we define $A(\mathcal{E};\omega)$ such that $A(\mathcal{E};\omega) d\omega d\mathcal{E}$ is the (joint) probability of photon absorption of frequency between $\omega$ and $\omega + d\omega$ for final energy $\mathcal{E}$ measured with respect to the bottom of the band at $\mathcal{E} = 0$. We define this probability as
\begin{equation}\label{eq:A}
A(\mathcal{E}_{final} = \mathcal{E};\omega) = \frac{n_A(\omega)}{N_A} \frac{D_J(\mathcal{E},\mathcal{E} - \hbar \omega) \rho_J(\mathcal{E},\mathcal{E} - \hbar \omega)}{\int
D_J(\mathcal{E},\mathcal{E} - \hbar \omega) \rho_J(\mathcal{E},\mathcal{E} - \hbar \omega) d\mathcal{E}},
\end{equation}
where $D_J(\e_{final},\e_{initial})$ is the squared magnitude of a transition matrix element for the electronic process $\e_{initial} \to \e_{final}$; Further, $\rho_J$ is the population-weighted density of pair states,
\begin{eqnarray}\label{eq:rho_J}
\rho_J(\mathcal{E}_{final},\mathcal{E}_{initial}) = \left[f(\e_{initial}) \rho_e(\mathcal{E}_{initial})\right] \left[(1 - f(\e_{final}) \rho_e(\e_{final})\right],
\end{eqnarray}
and $\rho_e = \frac{3 n_e}{2 \e_F} \sqrt{\frac{\e}{\e_F}}$ is the density of states of a free electron gas~\cite{Ashcroft-Mermin}, $n_e$ being the electron density. Finally, $n_A(\omega)$ is the number density of absorbed $\hbar \omega$ photons per unit time between $\omega$ and $\omega + d\omega$ and $N_A = \int d\omega n_A(\omega)$ is the total number density of absorbed photons per unit time. For CW illumination, it is given by
\begin{equation}\label{eq:N_A}
N_A = \frac{\langle p_{abs}\left(\vec{r},t\right)\rangle_t}{\hbar \omega},
\end{equation}
where the absorbed optical power density (in units of $W/m^3$) is given by the Poynting vector~\cite{Jackson-book}, namely,
\begin{equation}\label{eq:_P_abs}
\langle p_{abs}\left(t\right)\rangle_t = \omega \epsilon''(\omega,T_e,T_{ph}) \langle \vec{E}(t) \cdot \vec{E}(t)\rangle_t,
\end{equation}
where the temporal averaging, $\langle \rangle_t$, is performed over a single optical cycle such that only the time-independent component remains. Note that the absorption lineshape arises naturally from the spectral dependence of the local electric field in Eq.~(\ref{eq:_P_abs}); it depends on the nanostructure geometry and the permittivities of its constituents. This way, there is no need to introduce the lineshape phenomenologically as done in~\cite{Seidman-Nitzan-non-thermal-population-model}.

The absorption probability of a $\hbar \omega$ photon, $A$~(\ref{eq:A}), satisfies
\begin{equation}\label{eq:total_absorption}
\int_0^\infty A(\mathcal{E};\omega)d\mathcal{E} = \frac{n_A(\omega)}{N_A},
\end{equation}
and the net change of electronic population at energy $\mathcal{E}$ per unit time and energy at time $t$ due to absorption is $N_A \phi_A$, where
\begin{equation}\label{eq:phi_A}
\phi_A(\mathcal{E};\omega) = \int_0^\infty d\omega [A(\mathcal{E};\omega) - A(\mathcal{E} + \hbar \omega;\omega)],
\end{equation}
is a quantity describing the total (probability of a) population change at energy $\mathcal{E}$ per unit time and energy at time $t$. 

Altogether, the change of population due to photon excitation is given by
\begin{eqnarray}\label{eq:QM-like-excitation}
\left(\frac{\partial f}{\partial t}\right)_{ex}(\e) = \frac{N_A \phi_A(\e)}{\rho_e(\e)},
\end{eqnarray}
so that electron number conservation is ensured, $\int d\e \rho_e(\e) \left(\frac{\partial f}{\partial t}\right)_{ex}(\e) \sim \int d\e \phi_A(\e) = 0$.

The functional form of Eq.~(\ref{eq:QM-like-excitation}) is shown in Fig.~\ref{fig:BE_terms}(a) 
- one can see a roughly flat, $\hbar \omega$-wide region of positive rate above the Fermi energy, and a corresponding negative regime below the Fermi energy. In that regard, the incoherent, quantum-like, $|\vec{E}|^2$ excitation term reproduces the predictions of the photoelectric effect. 
The slight asymmetry originates from the density of states $\rho_e(\e)$~\footnote{This asymmetry may grow if the energy dependence of $D_J$ will be taken into account. }. Some earlier papers, e.g.,~\cite{Munday_hot_es} (and potentially, also~\cite{Manjavacas_Nordlander}\footnote{In that paper, a similar calculation was done, namely, of the ``hot'' electron excitation rate (rather than their density); however, the results were not shown on a logarithmic scale, hence, it is difficult to observe the similarity. }) used excitation rates similar to those of Eq.~(\ref{eq:QM-like-excitation}) to {\em qualitatively} describe the steady-state ``hot'' electron density. However, such a qualitative estimate is appropriate {\em only} in case all other terms in the underlying equation are energy-independent. Clearly, from Fig.~\ref{fig:BE_terms}, this is not generically the case. As explained in more detail in~\cite{Dubi-Sivan-Faraday}, this approach does not describe correctly the electron distribution near the Fermi energy, but it can describe the electron distribution correctly far from the Fermi energy (via multiplication by the $e-e$ collision time). Unfortunately, the former effect is orders of magnitude more important.

Note that in our approach, we effectively assume that momentum is conserved for all transitions. A more accurate description requires one to distinguish between the electron states according to their momentum, as done e.g., in~\cite{vallee_nonequilib_2003} for a continuum of electron states and in~\cite{GdA_hot_es,Govorov_ACS_phot_2017} for discretized electron states. 
However, it is worth noting in this context that the numerical results in~\cite{Govorov_ACS_phot_2017} show that when considering an ensemble of many nanoparticles with a variation in shape (up to 40\%), quantization effects nearly disappear even for a 2nm (spherical) particle. Indeed, the analytical result (red lines in Figs. 4 and 5 of~\cite{Govorov_ACS_phot_2017}) for the high-energy carrier generation rate, obtained by taking the continuum state limit, is very similar to the exact discrete calculation averaged over the particle sizes. This shows that neglecting the possibility of momentum mismatch (which is the effective meaning of avoiding the energy state quantization, as essentially done in our calculations) provides a rather tight upper limit estimate. Having said that, we bear in mind that quantization effects may still be relevant in highly regular nanoparticle distributions, ordered nanoparticle arrays or single nanoparticle experiments.

\subsubsection{The $e-ph$ collision term}\label{app:e-ph}
In~\cite{delFatti_nonequilib_2000}, the rate of change of $f$ due to $e-ph$ collisions is derived from the Bloch-Boltzmann-Peierls form~\cite{Ziman-book,PB_Allen_e_ph_scattering}$^,$\footnote{This expression does not include Umklapp collisions.}, giving
\begin{eqnarray}\label{eq:e-ph-integral-term}
\!\!\!\!\!\!\!\! \left(\frac{\partial f}{\partial t}\right)_{e-ph} &= - \frac{\mathcal{X}^2 \sqrt{m_{eff}^*}}{4 \pi \rho \sqrt{2\e}} \frac{1}{\hbar v_{ph}} \int_0^{\e_D} d\e_{ph} \e_{ph}^2 \big\{
f(\e) \left([1 - f(\e + \e_{ph})] n(\e_{ph}) + [1 - f(\e - \e_{ph})] [n(\e_{ph}) + 1]\right) \nn \\
&- [1 - f(\e)] \left[f(\e + \e_{ph}) [n(\e_{ph}) + 1] + f(\e - \e_{ph}) n(\e_{ph})\right] \big\}.
\end{eqnarray}
Here, $\mathcal{X} \sim 2 \e_F /3$ 
is the effective deformation potential~\cite{delFatti_nonequilib_2000}, $\rho$ is the material density 
and $m_{eff}^*
$ is the effective electron mass~\footnote{In our simulations, we used the values for these parameters as given in~\cite{delFatti_nonequilib_2000}. However, it should noted that the value they quote for $\rho$ might have involved a typo, which in turn, might have been adjusted via the value of $\mathcal{X}$. Either way, the overall value obtained for the cumulative $e-ph$ term ($G_{e-ph}$) is found to be in excellent agreement with the value computed in several other studies. }. For simplicity, we further assume that the phonon system is in equilibrium, so that $n(\e_{ph}) = n^T(\e_{ph}; T_{ph}) = \left(e^{\frac{\e_{ph}}{k_B T_{ph}}} - 1\right)^{-1}$ is the Bose-Einstein distribution function where $\e_{ph}$ is the phonon energy and $T_{ph}$ is the phonon temperature. Eq.~(\ref{eq:e-ph-integral-term}) relies on the Debye model~\footnote{This is justified for noble metals, such as Ag, where only acoustic phonons are present. Assuming that these phonon modes are distinct and excluding Umklapp processes, only the longitudinal phonon acoustic mode is coupled to the electron gas. }, namely, a linear dispersion relation for the phonons is assumed, $\e_{ph} = v_{ph} \hbar |q|$, where $v_{ph}$ is the speed of sound ($\cong 3650$m/s in Ag) and $q$ is the phonon momentum. Beyond the Debye energy, $\e_D = k_B T_D \cong 0.015$eV for Ag, the density of phonon states vanishes. Previous work emphasized the {\em in}sensitivity of the non-equilibrium dynamics to the phonon density of states and dispersion relations, thus, justifying the adoption of this simple model~\cite{delFatti_nonequilib_2000,Brown_PRB_2016} and the neglect of the phonon non-equilibrium. More advanced models that account also for the possible non-equilibrium of the lattice exist (see e.g., in~\cite{Italians_hot_es,Baranov_Kabanov_2014}) but are relatively rare.

The two terms associated with $f(\e + \e_{ph})$ describe phonon absorption, whereas the two terms associated with $f(\e - \e_{ph})$ describe phonon emission. Fig.~\ref{fig:BE_terms}(b) shows the energy dependence of these four different processes described by Eq.~(\ref{eq:e-ph-integral-term}) for $T_e - T_{ph} = 0.2^\circ$K, neglecting the small non-thermal part of the distribution (justified a-posteriory). For this temperature difference, an estimate based on the relaxation time approximation for $e-ph$ collisions allows us to relate the magnitude of each term ($\sim 10^{12}/sec$) to a collision rate of $\sim 10$fs, in accord with the value sometimes adopted within this context~\cite{Dressel-Gruner-book}. 
However, since these four processes compete with each other, the resulting total change of the distribution due to $e-ph$ collisions is several orders of magnitude slower. Overall, one can see that~(\ref{eq:e-ph-integral-term}) has a rather symmetric, $\sim \hbar \omega_D$-wide Lorentz-like lineshape. For $T_e > T_{ph}$, the rate is negative (positive) above (below) the Fermi energy, reflecting the higher likelihood of phonon emission processes, i.e., that energy is transferred from the electrons to the phonons. In order to see this more clearly, we can calculate the rate of energy transfer between the electrons and phonons by multiplying by $\e \rho_e(\e)$ and integrating over all electron energies. The resulting integral, defined as $W_{e-ph} \equiv - \int_0^\infty \e \rho_e(\e) \left(\frac{\partial f}{\partial t}\right)_{e-ph} d\e$, is hardly distinguishable from its thermal counterpart, $W_{e-ph}^T \equiv - \int_0^\infty \e \rho_e(\e) \left(\frac{\partial f^T}{\partial t}\right)_{e-ph} d\e$, which is usually represented by $G_{e-ph} \left(T_e - T_{ph}\right)$~\cite{PB_Allen_e_ph_scattering}. For $T_e > T_{ph}$, the factor $\e \rho_e(\e)$ weighs favourably the region above the Fermi energy, such that $W_{e-ph}$ and $W^T_{e-ph}$ are positive. In~\cite{Brown_PRB_2016}, an ab-initio, parameter-free derivation of the electron-phonon coupling coefficient based on density functional theory found $G_{e-ph} \sim 
3 \cdot 10^{16} W / m^3 K$ for Ag, in agreement with values found in previous works~\cite{el_lat_relaxation,PB_Allen_e_ph_scattering,delFatti_nonequilib_2000,Italians_hot_es,contribution_to_heat_cap_el_ph_inter,G_nonthermal_Hopkins_2015}, and with a negligible temperature-dependence, up to about $3000^\circ $K.

We note that our approach accounts for the mutual effect $e-e$ collisions have on $e-ph$ collisions~\cite{non_eq_model_Lagendijk}, since $e-ph$ collisions are treated by the $f$-dependent rate~(\ref{eq:e-ph-integral-term}) (rather than within the relaxation time approximation).

\subsubsection{The $e-e$ collision term}\label{app:tau_ee}


\subsubsection{The $e-e$ collision rate}
The rate of $e-e$ collisions near thermal equilibrium is usually slower than the $e-ph$ collision rate (order of picoseconds) since they involve only deviations from the independent electron approximation~\cite{Ashcroft-Mermin}. However, away from thermal equilibrium, the $e-e$ collision rates of high energy non-thermal electrons increase substantially and can become comparable to the $e-ph$ collision rate or even faster (see Fig.~\ref{fig:BE_terms}(c)). Specifically, by Landau's Fermi liquid Theory (FLT)~\cite{Quantum-Liquid-Coleman}, the (effective) $e-e$ collision rate is given by
\begin{equation}\label{eq:tau_ee_FLT}
\tau_{e-e}^{-1}(\e) = K \left[
\left(\pi k_B T_e\right)^2 + \left(\e - \e_F\right)^2\right], 
\end{equation}
where $K = {m_{eff}^*}^3/ 8 \pi^4 \hbar^6 W_{e-e}$ is the characteristic $e-e$ scattering constant that contains the angular-averaged scattering probability $W_{e-e}$ and the effective mass of the electron, $m_{eff}^*$; for Au and Ag, $K = 2 \cdot 10^{14}/eV^2\ s$~\cite{non_eq_model_Lagendijk}. Similar variations of this expressions within a continuum of states description were used e.g., in~\cite{delFatti_nonequilib_2000,vallee_nonequilib_2003,Italians_hot_es} in the context of ultrafast illumination. The more recent calculations of the $e-e$ collision rate within a discretized electron energy description, e.g., in~\cite{GdA_hot_es,hot_es_Atwater} retrieved this functional dependence. 
Experimental data obtained via two photon photo-emission measurements are found in excellent agreement with the Fermi liquid based expression~(\ref{eq:tau_ee_FLT}), see discussion in~\cite{Italians_hot_es}\footnote{It should be noted, however, that some earlier studies (e.g.,~\cite{non_eq_model_Lagendijk}) employed a different expression for $\tau_{e-e}$ which incorporates a strong asymmetry with respect to the Fermi energy, based on the famous expression derived in~\cite[Pines \& Nozieres]{Quantum-Liquid}. However, Coleman~\cite{Quantum-Liquid-Coleman} showed that the Pines \& Nozieres expression is, in fact, unsuitable for our purposes and that the symmetric parabolic dependence of the collision rate on the energy difference with respect to the Fermi energy (as in ~\cite{hot_es_Atwater,GdA_hot_es,Govorov_ACS_phot_2017}) is in fact the correct one. Indeed, the Pines \& Nozieres traces the collision dynamics of a single electron, rather than the relaxation dynamics of the distribution as a whole; in other words, it accounts for scattering of electrons from a certain electronic state $\e$, but ignores scattering {\em into} that energy state, a process which cancels out the dependence of the scattering rate on the Fermi function. }.

\subsubsection{Energy conserving relaxation time approximation}
Since $e-e$ collisions are elastic (and within the approximation adopted here, also isotropic)~\cite{Lundstrom-book}, we can adopt the relaxation time approximation for sufficiently small deviation from equilibrium, and write
\begin{eqnarray}\label{eq:RTA}
(\Delta_\tau f)_{e-e} = - \frac{f(\e,T_e,T_{ph}) - f^T(\e,T_e)}{\tau_{e-e}(\e)}.
\end{eqnarray}
However, we note that the regular $e-e$ term does not conserve the energy of the electron system as a whole (although it is supposed to, by the elastic nature of $e-e$ collisions). As a remedy, we introduce a term $\mathcal{F}_{e-e}(\e)$, defined by the condition
$\int_0^\infty \e \rho_e(\e) \left[(\Delta_\tau f)_{e-e} + \mathcal{F}_{e-e}(\e)\right] d\e = 0$. This additional term ensures that the electron energy, defined as $\U_e \equiv \int_0^\infty \e \rho_e(\e) f(\e,T_e,T_{ph}) d\e$, is conserved. Such a term is regularly included in Boltzmann models of fluid dynamics, where it is known as the Lorentz term~\cite{Hauge-Boltzmann-Lorentz}, but to our knowledge, was not employed in the context of illuminated metal nanostructures~\footnote{However, we note that in models that rely on the complete $e-e$ scattering integral (e.g., see examples in the context of ultrafast illumination~\cite{non_eq_model_Lagendijk,delFatti_nonequilib_2000,vallee_nonequilib_2003,Italians_hot_es}), the electron energy is conserved, so that the Lorentz term is not necessary. }. Thus, overall, we have
\begin{eqnarray}
\left(\frac{\partial f}{\partial t} \right)_{e-e} = (\Delta_\tau f)_{e-e} + \mathcal{F}_{e-e}(\e).
\end{eqnarray}
The absence of this term in previous steady-state derivations of the electron distribution (e.g., in~\cite{Govorov_1,Govorov_2,Govorov_ACS_phot_2017}) mean that energy is not conserved in these studies; Nevertheless, this specific effect is relatively small.

\subsubsection{Comparison between different scattering time functions}
\begin{figure}[h!]
\centering{\includegraphics[width=9cm]{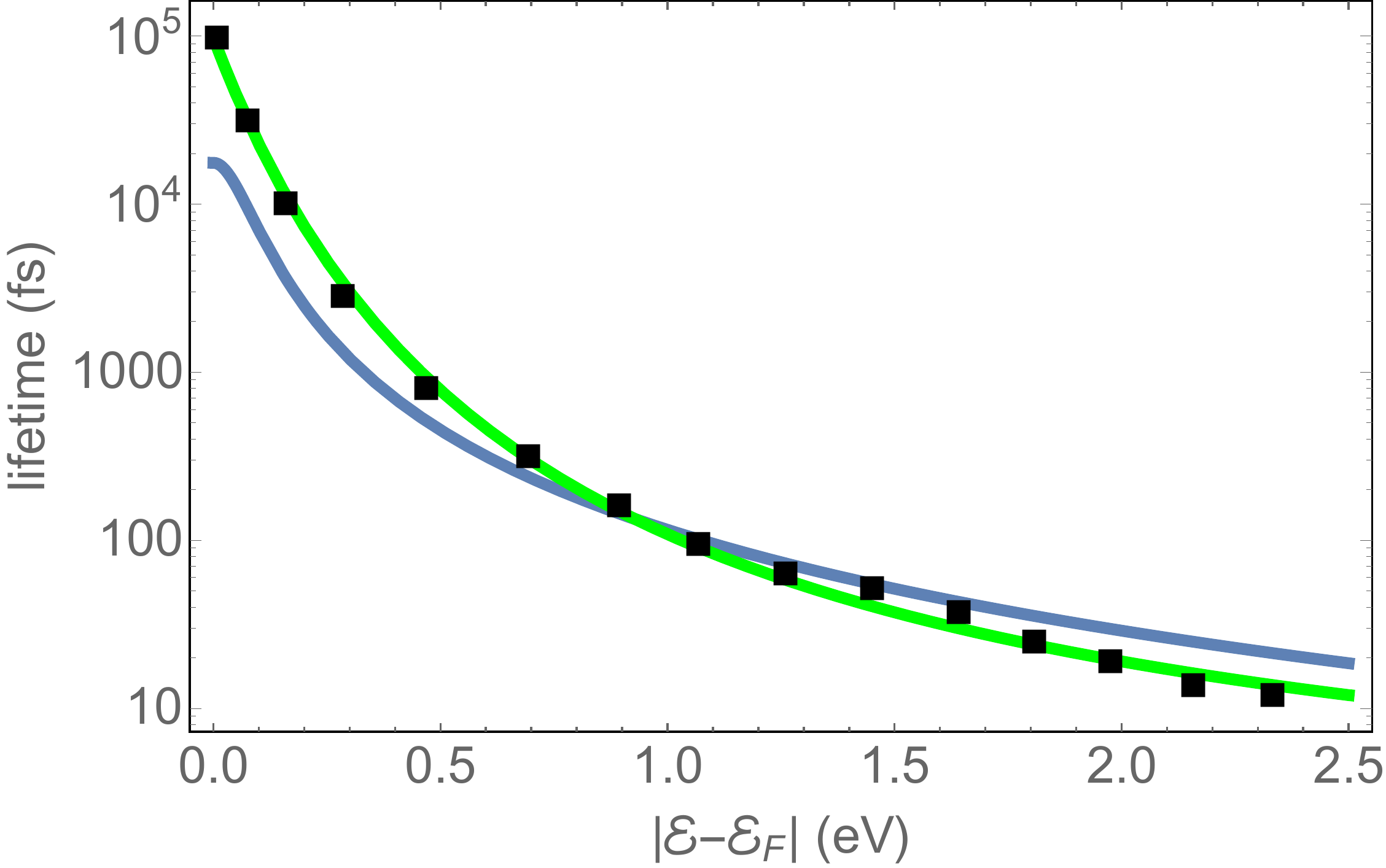}}
\caption{(Color online) The $e-e$ scattering time,  evaluated  by averaging over the data of from Ref.~\cite{GdA_hot_es} (black squares). Solid blue line is a fit to the standard (Fermi liquid) collision time~(\ref{eq:tau_ee_FLT}), and the solid green line are fits to the phenomenological form defined in the text.} \label{fig:Fig_S_abajo_fit}
\end{figure}

The results in the main text were obtained using an $e-e$ collision time of the form~(\ref{eq:tau_ee_FLT}). The coefficient $K$ was found using a fit to the calculations of Ref.~\cite{GdA_hot_es}. In addition, we performed the same calculation with a phenomenological scattering time of the form $\tau_{e-e}(\e) = e^{(1/(a + b \e}$), with a = 0.08585, b = 0.1278 eV$^{-1}$ (which decays slightly faster than the usual energy-dependence of the FLT); this seems to fit the data of Ref.~\cite{GdA_hot_es} better. In Fig.~\ref{fig:Fig_S_abajo_fit} we show the original data of Ref.~\cite[Fig. 6]{GdA_hot_es} and the fits to the standard FLT form~(\ref{eq:tau_ee_FLT}) and the phenomenological form.

\begin{figure}[h!]
\centering{\includegraphics[width=9cm]{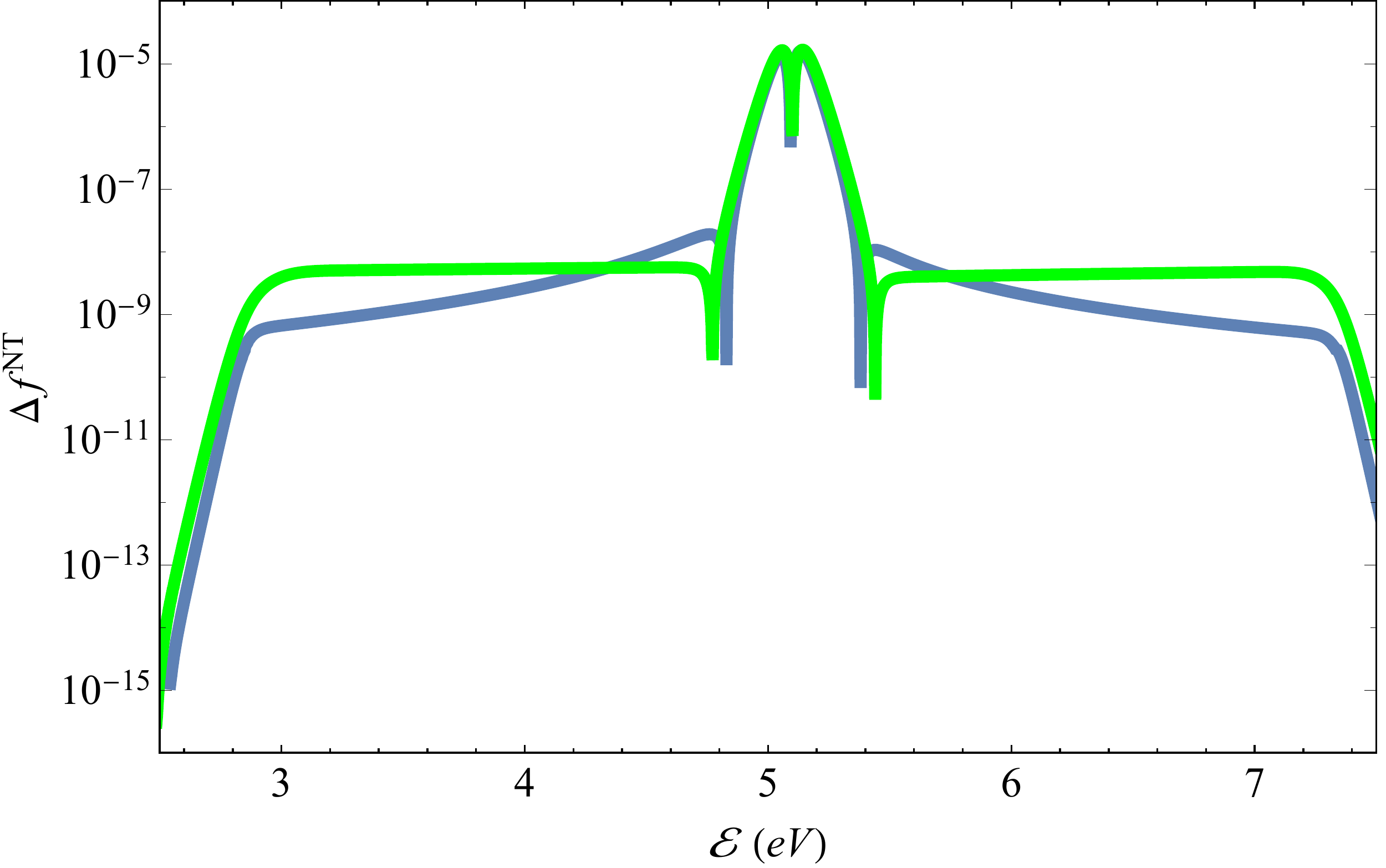}}
\caption{(Color online) Non-equilibrium electron distributions (see main text) for the two forms of $e-e$ scattering time, showing only a qualitative difference.} \label{fig:Fig_S_abajo_distribution}
\end{figure}

In Fig.~\ref{fig:Fig_S_abajo_distribution} we show the ``hot'' electron distribution evaluated with these two forms for the $e-e$ collision time, the FLT one and the phenomenological one. As can be seen, while the distributions are slightly different, the difference is essentially quantitative. The electron and phonon temperatures were found to be identical for the 2 expressions for $\tau_{e-e}$ (within our numerical accuracy for the 2 cases). 

\subsubsection{Evaluating the power density for different process}\label{sub:power}
The formalism presented in this manuscript allows us to evaluate the power density that goes into heating the electrons and phonons, and the power that goes into generating non-thermal carriers.

To evaluate this, we start with the general expression for the total energy of the electron system, $\mathcal{U}_e$ defined above.
Formally taking the time derivative gives the total power output of the electron system (which, at steady-state, vanishes by definition), \beq \label{power_tot}
\frac{d \mathcal{U}_e}{d t} = \int \e \rho_e(\e) \left(\frac{\partial f}{\partial t}\right) d\e.
\eeq
From Eq.~(\ref{eq:BE}), one can formally break $\left(\frac{\partial f}{\partial t}\right)$ into different contributions. Plugging these contributions into Eq.~(\ref{power_tot}) the power that goes into the difference energy channels. Specifically, substituting $\left(\frac{\partial f}{\partial t}\right)_{ex}$ gives the expression for the total power that is pumped into the electronic system by the photons.

Similarly, our formalism provides a natural way to distinguish between thermal and non-thermal contributions, since the steady-state distribution is naturally {\sl a-priori} defined as $f(\e) = f^T(\e,T_e) + f^{NT}(\e)$. The first term is a thermal distribution with the (elevated) steady-state electron temperature, and the second term is the non-thermal distribution. Thus, substituting these into the expression for power gives the power $W^T$ that goes into the thermal part of the electron distribution (i.e., that goes into electron heating) and the power $W^{NT}$ that goes into generating non-thermal carriers, namely,
\begin{eqnarray} \label{power_T}
W^T &=& \int \e \rho_e(\e) \left[\left(\frac{\partial f}{\partial t}\right)_{ex} \right]_{f=f^T} d\e, \\
W^{NT} &=& \int \e \rho_e(\e) \left[\left(\frac{\partial f}{\partial t}\right)_{ex} \right]_{f = f^{NT}} d\e.
\end{eqnarray}

\subsubsection{Electron tunneling from the nanoparticle}\label{app:tunnel}
The use of plasmonic naonparticles for applications requires that the ``hot'' electrons tunnel out of the nanoparticle in order to perform some function, be it tunneling into a molecular orbital for photocatalysis or across a Schottky barrier with a semiconductor for detection. The underlying assumption of much of the literature is that if such a process occurs at a given energy, then, the efficiency of the process will be proportional to the electron distribution at that energy. For example, in discussing tunneling across a barrier, then the eficiency of the process will be simply an integral over the electron distribution function over energies higher than the barrier energy (with some weight). Similarly, for tunneling into a molecular level, the efficiency will be proportional to the electron distribution at that energy.

In a recent paper~\cite{Dubi-Sivan-Faraday} we have shown that this is indeed the case for photo-catalysis, as long as the tunneling time is long compared to all other timescales, most importantly the $e-e$ scattering time. The argument relies on evaluating the distribution function in the presence of a tunneling term. We start by describing a tunneling term of the form $g(\varepsilon) f(\varepsilon)$, where $f$ is the distribution, and $g(\varepsilon)$ is some kernel, describing the tunneling rate per energy. Importantly, (i) $g$ is independent of the distribution and (ii) is localized at energies far from the Fermi energy, where the distribution is small (in fact, it could be a step-like function, for example if there is tunneling through a Schottky barrier, but here we have in mind photo-catalysis. The explanations below actually apply for both classes of applications).

Now, assuming that we know the steady-state distribution $f_0(\varepsilon)$, we look for a correction to it, $f = f_0 + f_1$. The next step is to linearize the bare Liouvillian (i.e. the right-hand-side of the Boltzmann equation without the tunnelling term). Then, for the steady-state we have
\begin{equation}\label{eq:approx_f_tunnel}
0 = \alpha(\varepsilon) f_1(\varepsilon) + g(\varepsilon)(f_0(\varepsilon) + f_1(\varepsilon)),
\end{equation}
where $\alpha(\varepsilon)$ is the linearization term. This equation can easily be solved to give $f_1 = \frac{g}{g + \alpha} f_0$. Now, as long as the dependence of $\alpha$ on energy is rather weak (which is indeed the case for both $e-e$ collision time and the excitation term), and the dependence of $f_0$ itself on $\varepsilon$ is also weak, the correction to the distribution function is simply proportional to the tunneling term $g(\varepsilon)$.

\begin{figure}
    \centering
    \includegraphics[width=0.7\textwidth,scale=0.1]{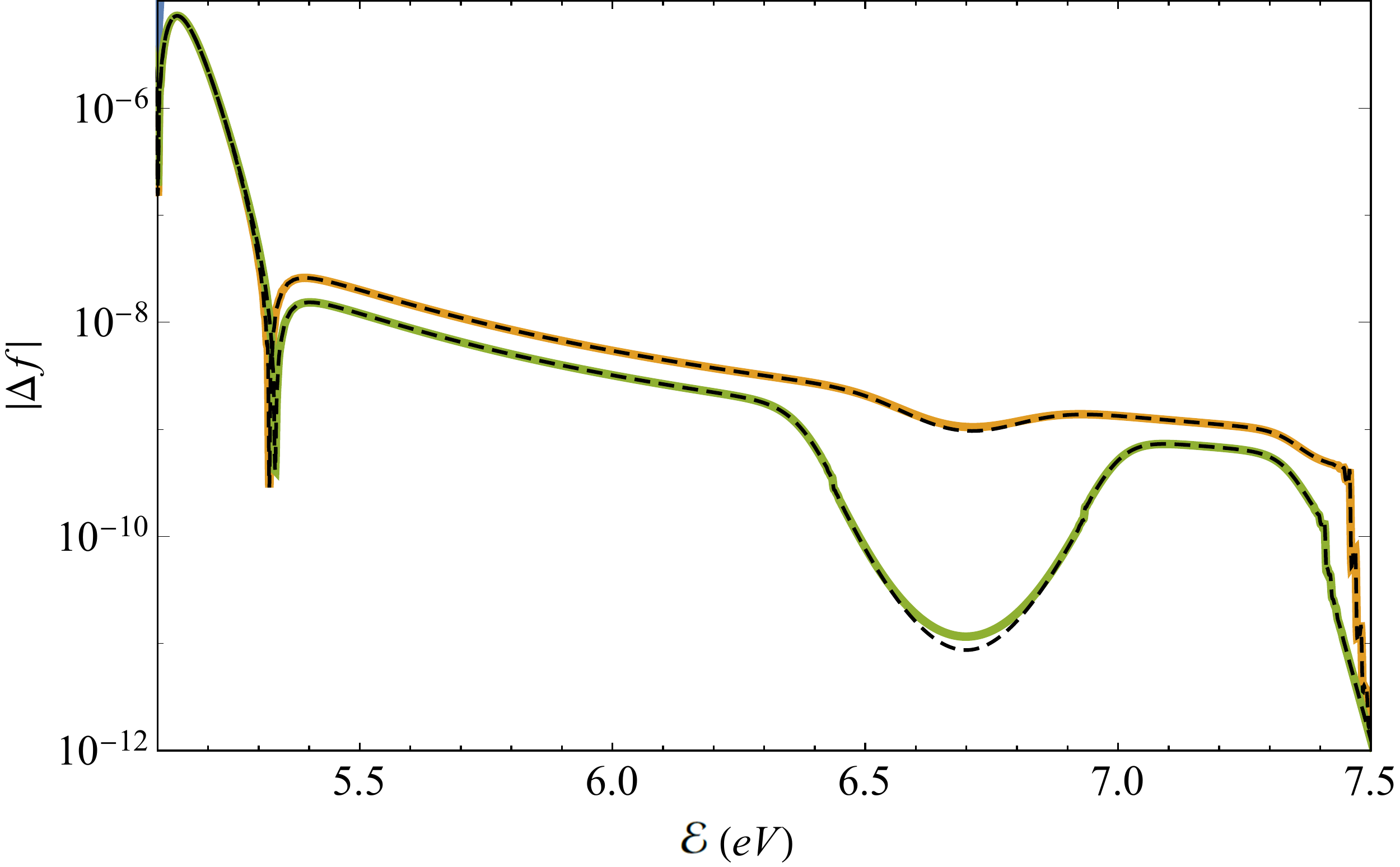}
    \caption{Non-equilibrium electron distribution with a tunneling term described in the text above, for $\gamma_T = 10^{13}$Hz (orange line) and $10^{15}$Hz (green line). The dashed lines are the approximations~(\ref{eq:approx_f_tunnel}). }
    \label{fig:tunnel}
\end{figure}

In order to test this (rather simple) estimate, we ran our calculation with an additional tunneling term of the form $-\gamma_T g(\varepsilon) f(\varepsilon)$, where $\gamma_T = 10^{13}$Hz and $10^{15}$Hz, corresponding to a slow ($100$ femtosecond) and fast (few femtosecond) tunneling time (which is extremely fast, as realistic tunneling times were shown to be as short as 100 fs only in the best case scenario, see e.g.,~\cite{Uriel_Schottky_2018}); $g$ is centered at $\approx 1.5$eV above the Fermi energy and has an energy width of a few hundreds of meV. In Fig.~\ref{fig:tunnel} we show the electron distribution with the tunneling terms, and the approximation~(\ref{eq:approx_f_tunnel}). As can be seen, for $\gamma_T = 10^{13}$Hz the approximation above is excellent.

Even more surprising and interesting, while for $\gamma_T = 10^{15}$Hz there should be a difference (because formally we are outside the regime of the approximation), still the approximation seems very good. The conclusion we draw from this calculation is that, in principle, and over a wide range of parameters (and physical processes), knowledge of the bare distribution function (i.e., evaluated without a tunneling terms) provides an excellent indication to the performance of the ``hot''-electron system as a functional device.

\subsection{Practical considerations}\label{app:practical}

In order to avoid limiting the generality of our results, we did not indicate throughout the manuscript details of a specific nanostructure. In this SI Section, we discuss what needs to be done in order to apply our theory to a specific experimental configuration. For simplicity, we discuss nanospheres; extension of the discussion to other particle shapes is possible.

\subsubsection{Local field}~\label{app:local_field}
Throughout the manuscript, we treated $|\vec{E}|$ as a parameter representing the {\em local} field~\footnote{Also note that throughout the manuscript we avoid specifying the local intensity, as it is a somewhat improper quantity to use when discussing metals. Indeed, the negative real part of the permittivity causes the fields within the metal to be primarily evanescent, hence, not to carry energy (such that the Poynting vector, hence, intensity vanish, at least in the absence of absorption). Instead, we use the local density of electromagnetic energy, by specifying the local electric field, which is easy to connect to the incoming field.}. In order to evaluate the non-thermal carrier density for an actual nanostructure configuration and illumination pattern, one needs to solve the Maxwell equations for the given configuration (for example, $\vec{E} = \left[3\epsilon_h / (2 \epsilon_h + \epsilon_m)\right] \vec{E}_{inc}$ for a small sphere illuminated uniformly) 
and apply our formulation {\em locally}, i.e., for each point in the nanostructure independently; this procedure was adopted in~\cite{Govorov_ACS_phot_2017} and was complemented by surface/volume averaging. In that respect, the role of surface plasmon resonances in promoting ``hot'' carrier generation is obvious - at resonance, the local electric fields are enhanced, hence, the electron system is driven more strongly away from equilibrium.

For weak electric fields, like used in the current work and essentially in all relevant experiments (see e.g.,~\cite{Halas_dissociation_H2_TiO2,Halas_H2_dissociation_SiO2}, the distribution and temperatures can then be readily determined. For small spherical metal nanoparticles, the temperature(s) are uniform~\cite{thermo-plasmonics-basics,Un-Sivan-size-thermal-effect}. The majority of previous theoretical studies relied on these same assumptions (e.g.,~\cite{Govorov_1,Manjavacas_Nordlander,GdA_hot_es}). 

For more complicated geometries, or for bigger nanostructures, the field may not be uniform. Nevertheless, the gradients of the electric fields are usually assumed to have a small effect on the electron distribution. The non-uniformity of the temperature is negligible, due to the relatively high thermal conductivity of the metal~\cite{thermo-plasmonics-basics,Un-Sivan-size-thermal-effect}. Due to these reasons, these gradients were neglected in all previous studies; we adopt the same approach here. For higher fields, the optical and thermal properties of the metal may change due to the rise in temperature, requiring a fully self-consistent solution of the coupled Maxwell, Boltzmann and heat equations. Such a treatment is left to a future study.

\subsubsection{Particle size}
The size of the particle affects the field relatively weakly for sufficiently small size (for which the quasi-static approximation holds). However, as well-known~\cite{thermo-plasmonics-basics}, the nanoparticle temperature depends strongly on the particle size; for example, for nano-spheres, it grows quadratically with the radius $a$. In our formulation, this effect is accounted for via the value of the phonon-environment coupling, $G_{ph-env}$, which is usually calculated from first principles via molecular dynamics simulations, see e.g.,~\cite{Kapitza-Au-silicon,Cahill-Kapitza-2006,Kapitza-NPs,Munjiza_2014}. Overall, it scales inversely with the surface area~\cite{vallee_nonequilib_2003} for nano-spheres; this is equivalent to assuming the total heat conductance to the environment is proportional to the particle surface area; this scaling facilitates estimates for non-spherical particles.

\begin{figure}[h!]
\centering{\includegraphics[width=16cm,height=5.6cm]{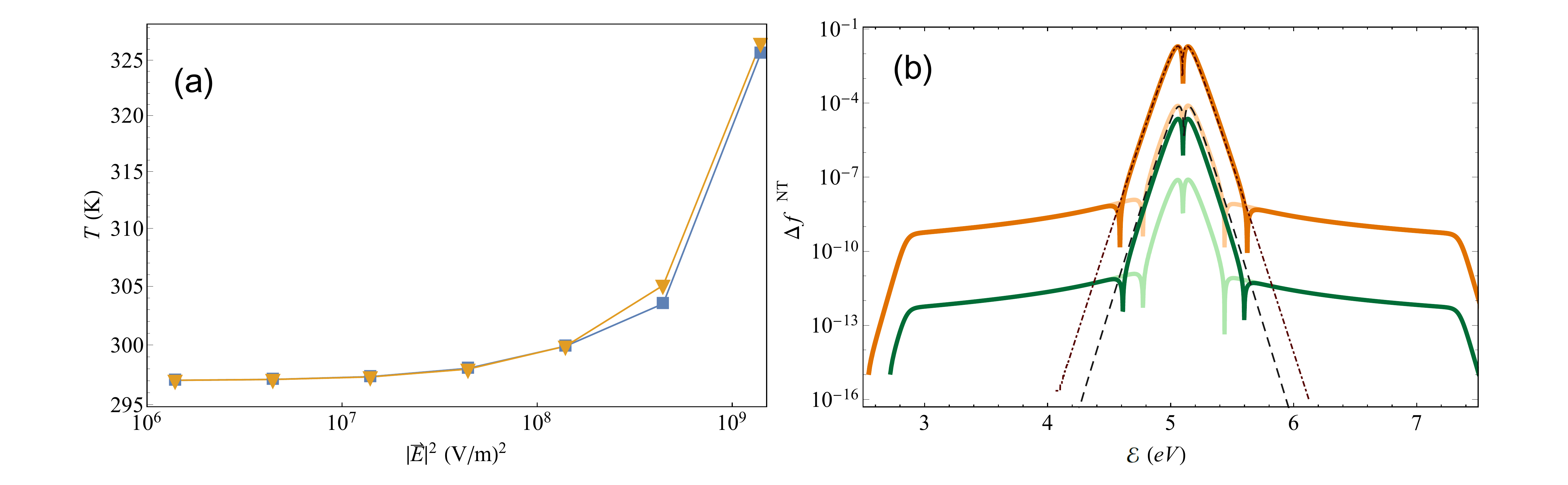}}
\caption{ (a) Electron (yellow) and phonon (blue) temperatures as a function of $|\vec{E}|^2$, for a system with a host thermal conductivity $G_{ph-env} = 5 \times 10^{12}$ W/m$^3$K, two orders of magnitude smaller than that employed in the simulations shown in Figs.~1-4 of the main text. Correspondingly, the temperature rise is much larger, as well as the difference between the electron and phonon temperatures. 
(b) Deviation of the non-equilibrium distribution from the thermal distribution for low host thermal conductivity and two intensities, $|\vec{E}|^2 = 1.4 \times 10^6,1.4 \times 10^8$(V/m)$^2$ (dark green and dark orange lines, respectively). For comparison, the distributions from the high $G_{ph-env}$ values used in Fig.~1 are also plotted (light green and light orange solid lines). The dashed black lines show the differences between simple Fermi functions with $T = 297.9$K (which is the electron temperature corresponding to the dark green line) and $T = 325$K (which is the electron temperature corresponding to the dark orange line) to a Fermi function at ambient temperature of $297$K.  
} \label{fig:SM2}
\end{figure}

As pointed out in the main text, we have carried out additional calculations to demonstrate the dependence of electron distribution and temperatures on the particle size via $G_{ph-env}$. The original results (appearing in the main text figures; $5$nm particle size; $G_{ph-env} = 5\times 10^{14}$ W/m$^3$K
~\cite{Kippelen_JAP_2010}) can now be compared to results for a particle which is 10 times bigger ($50$nm; $G_{ph-env} = 5 \times 10^{12}$ W/m$^3$K). In Fig.~\ref{fig:SM2}(a), we plot the electron and phonon temperatures as a function of intensity for these two cases. As can be observed, the electron temperature rise is $\sim 100$-fold larger for the larger particle (compare to Fig.~3), namely, about~30K. However, the difference between the electron and phonon temperatures is roughly the same; indeed, it can be shown analytically to be proportional to the incoming intensity which is the same for both sets of simulations.

In Fig.~\ref{fig:SM2}(b) we plot the electron non-equilibrium distribution (specifically, the absolute value of the deviation of the electron distribution from the Fermi distribution, $|\Delta f|$) for the two particle sizes 
and for two illumination levels, $|\vec{E}|^2 = 1.4 \times 10^6$, $1.4\times 10^9$(V/m)$^2$
. It is readily seen that the only deviations between the large and small particle cases are at the vicinity of the Fermi energy, but the non-thermal parts of the distributions (i.e., further away from $\e_F$, where $\Delta f^{NT} \sim \Delta f$) are {\em insensitive} to the particle size. In particular, we find that the efficiency of non-thermal high energy electron generation is independent of particle size, but the overall heating scales as $a^2$, in agreement with the single temperature (classical) heat equation. Such correspondence is absent in the simulations in~\cite{Govorov_ACS_phot_2017}
~\footnote{In~\cite{Govorov_ACS_phot_2017}, the electron temperature was not evaluated self-consistently, as in our formulation, but rather, it was set by hand and referred to as an ``effective'' temperature; no discussion of the choice of values was given. Unfortunately, the effective electron temperature values were set to $\approx 1300K$ ($0.1$ eV for a $4$nm NP), whereas the single temperature (classical) calculation for this configuration shows that the temperature rise should be $\approx 0.13K$. In addition, the scaling of the effective temperature used in~\cite{Govorov_ACS_phot_2017} violates the classical $a^2$ scaling; in fact, it showed an inverse proportionality to the NP size (specifically, the effective temperature of a 24nm NP was $\approx 520$K ($0.04$eV)). Claims in~[Govorov \& Besteiro, ArXiv 2019] on the emergence of quantum effects in this context are questionable, due the relatively large size of the NPs studied in this case, see also the discussion at the end of Section~\ref{app:QM_absorption}. }. This also means that smaller particles give rise to a higher relative efficiency of non-thermal carrier generation. This prediction should motivate a careful, single particle study that will enable one to verify this prediction vs. potentially contradicting claims based on measurements from macroscopic nanoparticle suspensions.

The results of Fig.~\ref{fig:SM2}(b) can be also interpreted in terms of the dependence of the non-thermal distribution on the host thermal conductivity. Indeed, the rate of energy density transfer to the environment $G_{ph-env}$ is also proportional to the thermal conductivity of the host~\cite{vallee_nonequilib_2003}. Thus, the different curves in Fig.~\ref{fig:SM2}(a) can be also associated with a system with a host thermal conductivity which is two orders of magnitude lower than the one presented in the main text. As for the larger nanoparticle, the electron temperature rise and the difference between the electron and phonon temperatures, are higher, as expected - indeed, the heat flows away from the nanoparticle much more slowly for the larger nanoparticle. This shows, as stated in the main text, that if ``hot'' electrons play a dominant role in some experiment (e.g. in photo-catalysis), then, the experimental results should be unaffected by a change of host. Conversely, if the results {\em are} affected by a change of host material (as observed e.g., in~\cite{Halas_dissociation_H2_TiO2,Halas_H2_dissociation_SiO2}), then it is not likely that the reason for that is the number of ``hot'' electrons, but rather due to a thermal effect, or an altogether different chemical effect; for a detailed discussion, see also~\cite{Y2-eppur-si-riscalda}.

\subsubsection{Surface scattering and quantum size effects} If one is interested in even smaller nanoparticles, then, within the energy state continuum description used in the current work, 
it may be necessary to account also for $e-surface$ collisions (the so-called ``quantum size effects''), as noted as early as in~\cite{Kreibig-book}. As this effect does not involve conservation of electron momentum, it can be accounted for in our formulation by adding a relaxation time like term, $(f - f^T)/\tau$, where $\tau$ is the time scale for these collisions which can be as fast as a few hundreds of femtoseconds in the case of a metal surface with atomic roughness~\cite{Uriel_Schottky_2018}; accordingly, it is practically negligible with respect to the $e-e$ and $e-ph$ collision rates. Depending on the nature of the $e-surface$ collisions, one may want to include/exclude them from the conservative term, $\mathcal{F}_{e-e}(\e)$.

However, it should be noted that in more advanced models where the energy states are discretized (such that $e-surface$ collisions are accounted for inherently), e.g.,~\cite{GdA_hot_es,Govorov_ACS_phot_2017}, the electronic states and the phononic states are extended throughout the bulk, and no ``surface states'' appear. One thus expects that in such calculations there will be no separate contribution from $e-surface$ collisions. In fact, in~\cite{GdA_hot_es,hot_es_Atwater} it was shown that the electron collision time is independent of the nanoparticle size. All these results indicate that unlike previous claims~\cite{hot_es_review_2015} ``quantum size effects'' have at most a small quantitative effect on the non-thermal carrier generation efficiency. This result was corroborated in~\cite{Govorov_ACS_phot_2017}, see discussion at the end of Section~\ref{app:QM_absorption}.

\end{document}